\documentclass[traditabstract]{aa}

\usepackage{graphicx,threeparttablex,longtable,array} 
\usepackage{amssymb,amsmath,multirow,gensymb,mathrsfs}
\usepackage{enumitem,multirow,rotating}
\usepackage[caption=false]{subfig} 
\usepackage{natbib}
\usepackage{ragged2e,epstopdf}
\usepackage[dvipsnames]{xcolor}
\usepackage{soul}
\usepackage[caption=false]{subfig}
\usepackage[left]{lineno}
\newcommand{\Jybeam}{\mbox{Jy beam$^{-1}$}}
\newcommand{\mJybeam}{\mbox{mJy beam$^{-1}$}}
\newcommand{\uJybeam}{\mbox{$\mu$Jy beam$^{-1}$}}
\newcommand{\kms}{\mbox{km s$^{-1}$}}
\newcommand{\um}{\mbox{$\mu$m}}

\newcommand{\Msun}{\mbox{M$_{\odot}$}}

\newcommand{\COO}{\mbox{C$^{18}$O}}
\newcommand{\HCCOp}{\mbox{H$^{13}$CO$^+$}}
\newcommand{\HCCN}{\mbox{H$^{13}$CN}}
\newcommand{\COrare}{\mbox{C$^{17}$O}}

\begin{document}

\title{Constraining the stellar masses and origin of the protostellar VLA~1623 system}

\author{Sarah I. Sadavoy\inst{1} \and
Patrick Sheehan\inst{2} \and
John J. Tobin\inst{2} \and
Nadia M. Murillo\inst{3,4} \and
Richard Teague\inst{5} \and
Ian W. Stephens\inst{6,7} \and
Thomas Henning\inst{8} \and
Philip C. Myers\inst{7} \and
Edwin A. Bergin\inst{9}
         }
         
\institute{Department of Physics, Engineering Physics, and Astronomy, Queen's University, Kingston, Ontario, K7L 3N6, Canada \and
National Radio Astronomy Observatory, Charlottesville, VA 22903, USA \and
Instituto de Astronom\'{i}a, Universidad Nacional Aut\'{o}noma de M\'{e}xico, AP106, Ensenada CP 22830, B. C., M\'{e}xico \and
Star and Planet Formation Laboratory, RIKEN Cluster for Pioneering Research, Wako, Saitama 351-0198, Japan \and
Department of Earth, Atmospheric, and Planetary Sciences, Massachusetts Institute of Technology, Cambridge, MA 02139, USA \and
Department of Earth, Environment, and Physics, Worcester State University, Worcester, MA 01602, USA \and
Harvard-Smithsonian Center for Astrophysics, 60 Garden Street, Cambridge, MA 02138, USA \and
Max-Planck-Institut f\"{u}r Astronomie, K\"{o}nigstuhl 17, D-69117 Heidelberg, Germany \and
Dept. of Astronomy, University of Michigan, Ann Arbor, MI 48104, USA
}


\date{Received ; accepted}

\abstract{We present ALMA Band 7 molecular line observations of the protostars within the VLA 1623 system.  We detect \COrare~(3 -- 2) in the circumbinary disk around VLA 1623A and the outflow cavity walls of the collimated outflow.  We further detect redshifted and blueshifted velocity gradients in the circumstellar disks around VLA 1623B and VLA 1623W that are consistent with Keplerian rotation. We used the radiative transfer modelling code \texttt{pdspy} and simple flared disk models to measure stellar masses of $0.27 \pm 0.03$ \Msun, $1.9^{+0.3}_{-0.2}$ \Msun, and $0.64 \pm 0.06$ \Msun\ for the VLA~1623A binary, VLA~1623B, and VLA~1623W, respectively.   These results represent the strongest constraints yet on stellar mass for both VLA~1623B and VLA~1623W, and the first mass measurement for all stellar components using the same tracer and methodology.  We use these masses to discuss the relationship between the young stellar objects (YSOs) in the VLA 1623 system.  We find that VLA~1623W is unlikely to be an ejected YSO, as has been previously proposed. While we cannot rule out that VLA~1623W is a  unrelated YSO, we propose that it is a true companion star to the VLA~1623A/B system and that these stars formed in situ through turbulent fragmentation and have had only some dynamical interactions since their inception.  
}

\keywords{Stars: formation, Stars: protostars, Stars: low-mass, Protoplanetary disks, Binaries: general, Molecular data}

\authorrunning{Sadavoy et al.}
\maketitle

\section{Introduction\label{Intro}}

Stellar mass is perhaps the most fundamental stellar property. Nevertheless, measuring stellar masses for the  youngest protostars remains difficult because these objects are deeply embedded in an infalling core and envelope and cannot be observed directly in optical or near-infrared wavelengths \citep[e.g.][]{difran07,Tychoniec21}. Detections of Keplerian rotation in the protostellar disks around protostars are paramount for constraining the stellar masses of these systems. Historically, Keplerian rotation has been difficult to detect (1) because bright molecular lines such as CO tend to become optically thick within the envelope, obscuring the disk, and (2) due to a lack of high-spatial resolutions needed to disentangle the disk from the envelope \citep[e.g.][]{Ohashi97, Schreyer06, Jorgensen07prosac, Murillo13,vantHoff18}. 

The Atacama Large Millimeter/submillimeter Array (ALMA) can provide the high-resolution molecular line observations that are necessary to make detections of Keplerian rotation in young disks possible.  Moreover, the significant improvements in sensitivity with ALMA have opened the possibility of observing fainter and optically thin isotopologues, which allows us to detect disks through the envelope \citep[e.g.][]{Murillo13, Yen14,Ohashi14, Aso15,Ginsburg18,Reynolds21}. ALMA is even capable of exploring deviations from pure Keplerian rotation that could be due to planets \citep[e.g.][]{Pinte18, Pinte20, Teague19,Teague21}, though primarily in older protoplanetary disks where the envelope is less of a confounding factor.   Nevertheless, despite the improvements afforded by ALMA, well-constrained stellar masses have only been measured for a limited number of low-mass protostars to date \citep[e.g.][]{Murillo13,Ohashi14,Aso17,Aso23,Ohashi23,Flores23,Thieme23}.    

VLA~1623 is in the $\rho$ Oph A star-forming region at 140 pc \citep{OrtizLeon18} and is one of the rare protostellar systems with detected Keplerian rotation.  Specifically, VLA~1623 is the canonical Class 0 protostellar object \citep{Andre93} and a hierarchical multiple system, with sources VLA~1623A and VLA~1623B (separated by $\sim$ 140 au) located towards the centre of the core and VLA~1623W roughly 1500 au west of the pair\footnote{An additional source, VLA~1623NE, is 2700 au north-east of the central pair, but we exclude this source as a more evolved, unrelated Class II object \citep{Sadavoy19}.} \citep{BontempsAndre97,Looney00,MurilloLai13}.  VLA~1623A is itself a compact binary, with its Aa and Ab components separated by $\sim 14$ au \citep{Harris18} and surrounded by a large ($R \approx 180$ au) circumbinary disk \citep{Murillo13}.  Molecular line observations of the circumbinary disk show evidence of Keplerian rotation and yield a combined stellar mass of between 0.2 and 0.5 \Msun\ \citep{Murillo13, Hsieh20}.  Measuring the masses of the other stellar components, however, has been challenging due to confusion with the circumbinary disk and insufficient sensitivity to robustly detect and resolve the disks in gas tracers.

Constraining the stellar masses is an important step towards examining the origins of the VLA~1623 system.   Observations suggest that the B and W components may be more evolved (non-coeval) due to a lack of outflow and the shape of their spectral energy distributions (SEDs), which imply a later evolutionary stage \citep{Murillo18}.  \citet{Harris18} examined the proper motion of the stars and propose that VLA~1623W may have been ejected.  An ejection scenario could explain the lack of envelope detected towards VLA~1623W \citep{Michel22} as well as the large velocity streamers in the system \citep{Hsieh20,Mercimek23}. Conversely, previous molecular line observations have been used to estimate the other stellar masses \citep{Ohashi22, Mercimek23}, and those estimations, even if not concrete measurements, indicate that VLA~1623B and VLA~1623W may be relatively massive compared with VLA~1623A. If confirmed, these masses would pose difficulties for the ejection scenario \citep{Reipurth10}.

Here we present new molecular line observations of \COrare\ (3 -- 2) of the VLA~1623 system from ALMA.  We used these data and radiative transfer models of rotating disks to constrain the stellar masses for the stellar components in VLA~1623.  \COrare\ is a rarer species that is less optically thick within the envelope than the line tracers that have previously been observed, thereby making it easier for us to disentangle the Keplerian rotating disks of the protostars from each other. The paper is organized as follows: Section \ref{data} details the observations, Section \ref{results} shows that the \COrare\ traces the disk and the other lines trace outflows or infall, Section \ref{modeling} describes the radiative transfer models, Section \ref{discussion} discusses the stellar masses and the possible origins of the VLA~1623 system, and finally Section \ref{summary} gives our conclusions.


\section{Data}\label{data}

ALMA observed VLA 1623 in Band 7 on 20 and 22 April 2019 for project 2018.1.01089.S in the C-4 configuration with baselines of 15 m -- 740 m.  The phase calibrator was J1650-2943, and the flux/bandpass calibrator was J1924-2914 (20 April) and J1517-2422 (22 April).  The data phase centre was at 16:26:26.35 -24:24:30.55, between VLA~1623A and VLA~1623B, meaning that VLA~1623W is outside of the primary beam full width at half maximum (FWHM)\footnote{The primary beam response is 0.4 at the position of VLA~1623W.}.  The time on source was 52.74 minutes in total.  The observations were set up with spectral windows on \COrare~(3 -- 2) at 337.061 GHz, \HCCOp~(4 -- 3) at 346.998 GHz, SO$_2$~($8_{2,6}-7_{4,4}$) at 334.673 GHz, and \HCCN~(4 -- 3) and SO$_2$~($13_{2,12}-12_{1,11}$) at 345.34 GHz.   There was also one continuum window with a central frequency of 348.2227 GHz and bandwidth of 1.85 GHz.  

We applied self-calibration to the continuum window and the three line windows using line-free channels.  We used three rounds of phase-only self-calibration, starting with a shallow clean and a long solution interval  and progressing to deeper cleans with shorter solution intervals.  Due to the high ($> 1000$) source signal-to-noise ratio (S/N), we also used two rounds of phase and amplitude self-calibration using long integration times. We used solution intervals of 3 min, 60.6 s, and 30.3 s for the phase-only self-calibration and 3 min and 90.9 s for the amplitude and phase self-calibration. The clean depths for the five rounds were 0.92, 0.45, 0.20, 0.17, and 0.13 \mJybeam.  The final continuum map including all line-free channels have a sensitivity of 95 \uJybeam\ at 341.8541~GHz for a beam of $0.3\arcsec \times 0.24\arcsec$\ with Briggs weighting and $\mbox{\texttt{robust}}=-0.5$. The maximum recoverable scale for the data is 3.8\arcsec\ using the 5th percentile baseline.

\indent The self-calibration gain solutions were applied to each of the line spectral windows.  We applied continuum subtraction using a fit order of 1.  The final line cubes are made using \texttt{tclean} interactively with $\mbox{\texttt{robust}}=0.5$. Table \ref{line_info} gives the rest frequency, channel width, sensitivity per channel, and beam size for each detected line. For \COrare\ (3 -- 2), we report the rest frequency of the main hyperfine component only. The \HCCN\ (4 -- 3) and SO$_2$ (13$_{2,12}$ -- 12$_{1,11}$) lines may be blended.  Hereafter, we refer to these data as (blended) \HCCN\ to distinguish these data from the other detected lines. Table \ref{line_info} also lists the velocity range used to make all moment maps (unless a different range is specified).

\begin{table*}[h!]
\caption{Line results.}\label{line_info}
\begin{tabular}{llllll}
\hline
Transition          & $\nu$ (GHz) & $\Delta v^a$ (km/s) & $\sigma$ (\mJybeam\ per chan)& beam ($a \times b, \theta$)   & $v_{range}^b$(km/s) \\
\hline\hline
SO$_2$ ($8_{2,6}-7_{4,4}$) & 334.673   & 0.219          & 3.1               & $0.413\arcsec \times 0.336\arcsec, -84\degree$ & $2.8 - 4.8$\\
\COrare\ (3 -- 2)    & 337.061       &  0.109            & 3.1               & $0.415\arcsec \times 0.337\arcsec, -87\degree$ &  $1.0 - 6.5$\\
SO$_2$ ($13_{2,12}-12_{1,11}$)$^c$ & 345.338     &  0.106            & 2.9               & $0.401\arcsec \times 0.329\arcsec, -86\degree$ & $2.3 - 5.5$\\
\HCCN (4 -- 3)$^c$       & 345.340  & $\cdots$ & $\cdots$ & $\cdots$ & $\cdots$\\
\HCCOp (4 -- 3)      & 346.998       &  0.106            & 3.1               & $0.402\arcsec \times 0.327\arcsec, -85\degree$ & $2.4 - 5.8$ \\
\hline
\end{tabular}
\begin{tablenotes}[normal,flushleft]
\item $^a$ Channel width. 
\item $^b$ Range of velocities used for the moment maps in Figure \ref{mom0}.
\item $^c$ These two lines are blended.
\end{tablenotes}
\end{table*}


\section{Results}\label{results}

\subsection{Overview}

\begin{figure}[h!]
\includegraphics[width=0.485\textwidth]{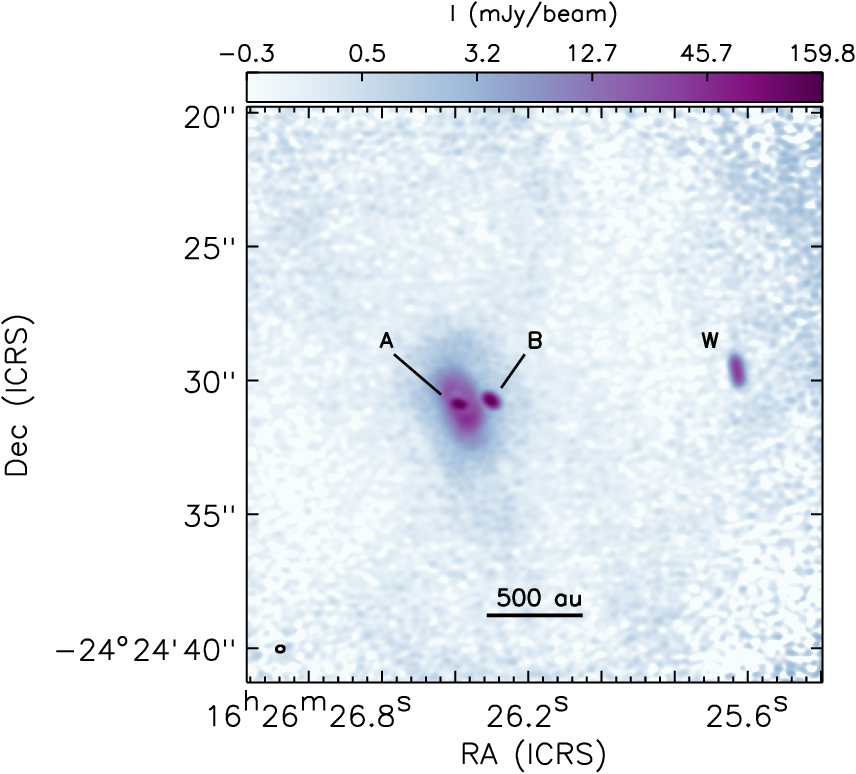}
\caption{Dust continuum map at 341.854 GHz (0.877 mm) zoomed in on the VLA 1623 sources.  Protostars are labelled A (for the unresolved Aa and Ab binary), B, and W. }   \label{b7_cont}
\end{figure}

Figure \ref{b7_cont} shows the dust continuum data for VLA 1623 with the main components labelled.  We recover the circumbinary disk around the protobinary, Aa and Ab, but we do not resolve the protobinary itself (labelled as A).  Hereafter, we refer to the protobinary as VLA~1623A. The circumstellar disks for B and W are also labelled.  

\begin{figure*}[h!]
\includegraphics[width=0.95\textwidth]{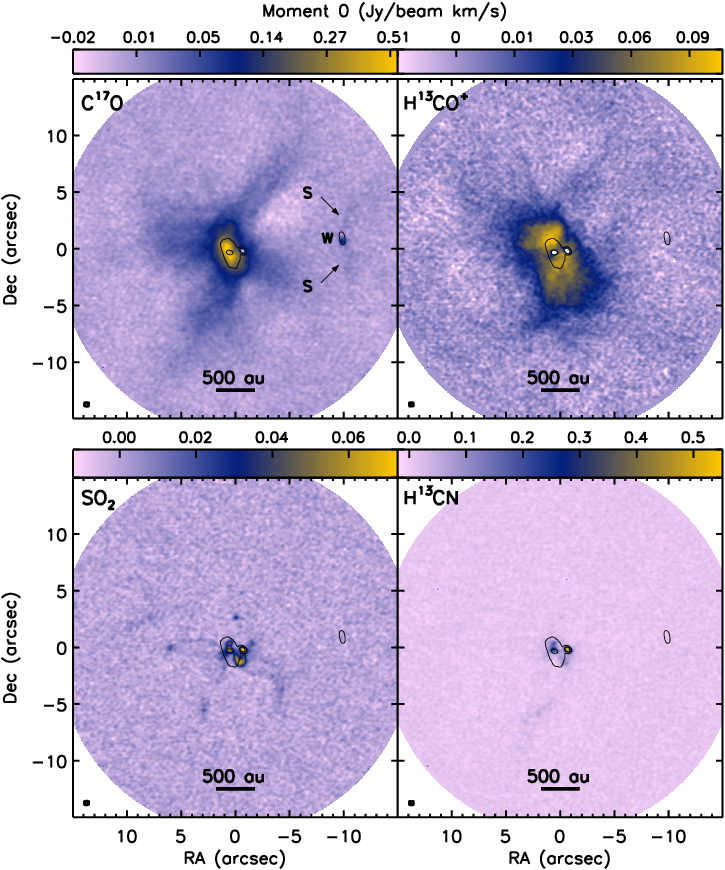}
\caption{Moment 0 maps (in Jy/beam km/s) of \COrare\ (3-2), \HCCOp\ (4-3), SO$_2$ (8-7), and (blended) \HCCN\ (4-3).  The velocity ranges are given in Table \ref{line_info}. 
Black contours show the Stokes I continuum at 7  and 49~\mJybeam.  The beam for each map is in the lower-left corner.  The upper panels use log scaling to highlight the extended emission, whereas the bottom panels have linear scaling.  The line maps are not primary-beam-corrected, for the sake of visualization. The first panel shows labels for VLA~1623W (W) and the velocity streamers (S) from \citet{Mercimek23} for clarity.}   \label{mom0}
\end{figure*}

Figure \ref{mom0} shows the Moment 0 data for each line with contours of dust continuum.  For the sake of clarity, the Moment 0 maps have not been primary-beam-corrected.  \COrare\ and \HCCOp\ have extended emission, where both molecules trace the circumbinary disk and inner envelope around A and B.   At the locations of the circumstellar disks for A and B, we see a relative deficit (absorption) in \COrare\ and \HCCOp\ that is consistent with the circumstellar disks shadowing the brighter circumbinary disk material or self-absorption \citep[e.g.][]{Murillo13, Hara21}.  The circumstellar material is bright in SO$_2$ and (blended) \HCCN, and both lines also show compact emission within the circumbinary disk towards the south-west.

Finally, although faint, we also find hints ($\sim 2.5\sigma$) of the vertical streamer passing through VLA~1623W in \COrare\ (3 -- 2) that matches what \citet{Mercimek23} found in \COO\ (2 -- 1) observations (labelled S in Figure \ref{mom0}).  VLA~1623W is only detected in \COrare\ (see Section \ref{circumdisk}), with a system velocity of about 1.8 \kms.  The system velocities of VLA~1623A and VLA~1623B are roughly 4 \kms\ and 2.3 \kms, respectively.  

\subsection{Outflows and infall}

\begin{figure}[h!]
\includegraphics[width=0.485\textwidth]{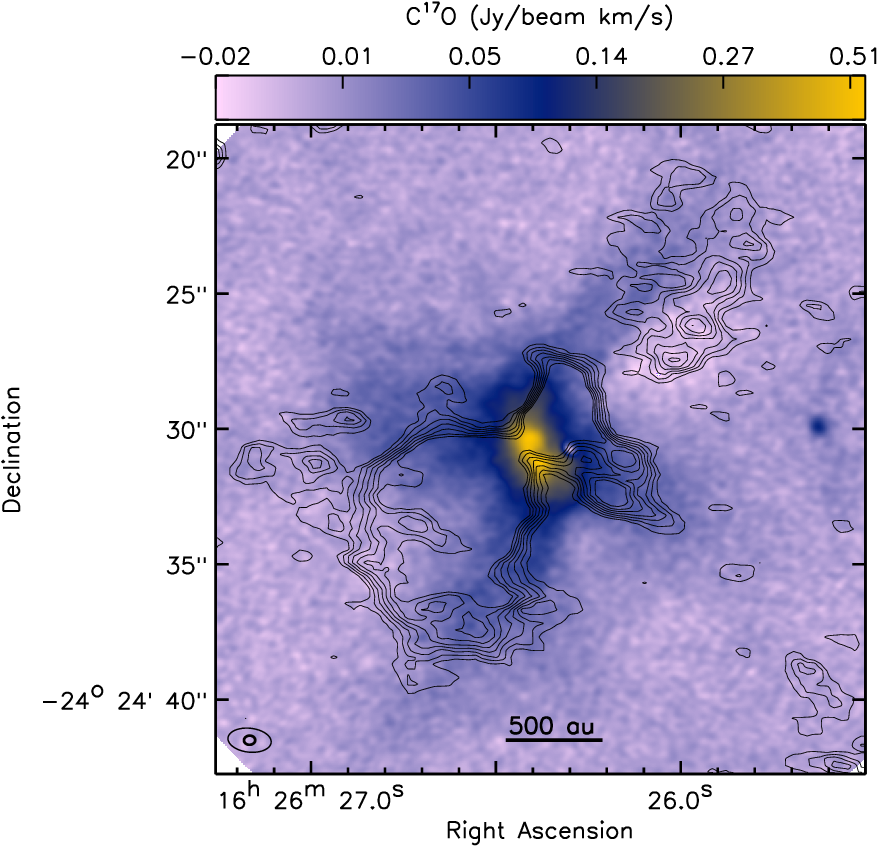}
\caption{\COrare\ (3 -- 2) data from Figure \ref{mom0} with contours of c-C$_3$H$_2$ (217.8 GHz) integrated intensity from \citet{Murillo18gas} to show the outflow.  The c-C$_3$H$_2$ integrated intensity data are evaluated over a velocity range of $2.65-5.36$ \kms, and the contours go from 0.02 to 0.05 \Jybeam\ \kms\ in steps of 0.005 \Jybeam\ \kms.  The map resolutions are given in the bottom-left corner. }   \label{cavitywall}
\end{figure}

The \COrare\ data show a large X-shaped pattern centred on the VLA~1623A/B system (see Figure \ref{mom0}).  Weaker emission along this pattern is also seen in SO$_2$ and (blended) \HCCN, indicating that this emission may be associated with the outflow and not the inner envelope. VLA~1623A has overlapping outflows with position angles of $\sim 125$\degree\ \citep[e.g.][]{Murillo18gas, Hara21, Ohashi22} in good agreement with the orientation of the extended \COrare\ data.  We find no \COrare\ gas within the outflow itself, however, suggesting that \COrare\ is tracing the outflow cavity walls only, likely due to higher densities in the walls over the outflow itself and higher temperatures leading to more sublimation of CO isotopologues. We note that \citet{Hsieh20} instead identify similar structures in \COO\ (2 -- 1) as gas streams in the envelope. To distinguish between these cases, Figure~\ref{cavitywall} compares the \COrare\ integrated intensity data with contours of c-C$_3$H$_2$ at 217.8 GHz from \citet{Murillo18gas}.  Hydrocarbons like c-C$_3$H$_2$ are excellent tracers of outflow cavities  \citep[e.g.][]{Sakai14,Tychoniec21,Ohashi22}.  Broadly, we find good agreement between c-C$_3$H$_2$ and the X-shaped extended emission in \COrare, which is why we attribute this material to the outflow wall rather than a gas streamer.   

\begin{figure}[h!]
\includegraphics[width=0.485\textwidth]{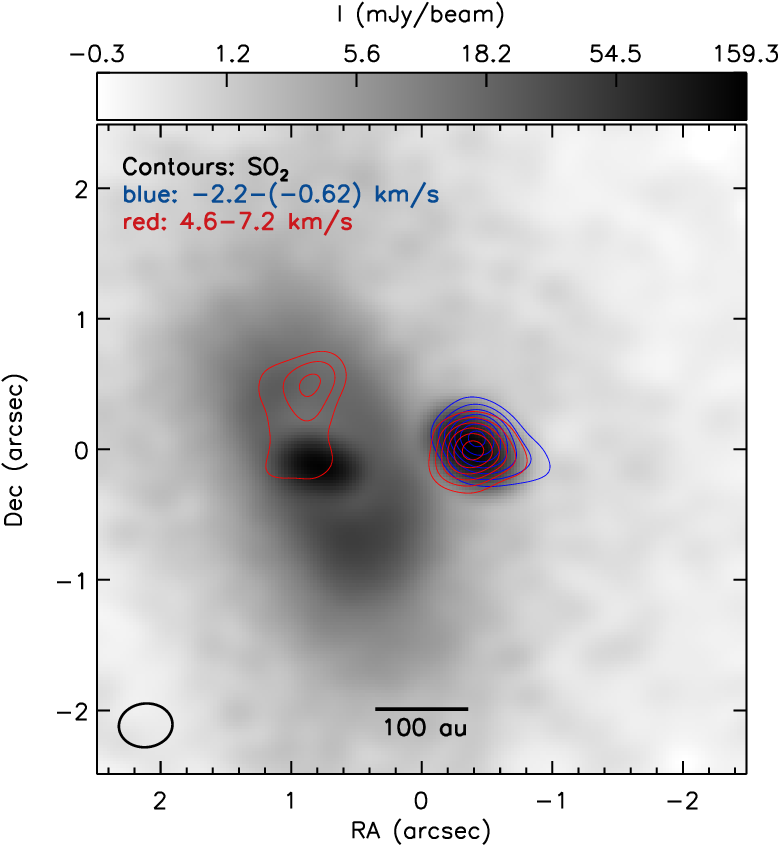}
\caption{Red- and blueshifted SO$_2$ (8 -- 7) emission overlaid on the dust continuum data.  The integrated intensities were generated using the velocity ranges indicated in the legend. Contours correspond to 10, 15, 20, 25, 30, 35, and 40 $\sigma$, where $\sigma = 2.5$ \mJybeam\ \kms\ for both the blue- and redshifted data. The SO$_2$ map resolution is in the lower-left corner. \label{gradientBso2}}
\end{figure}

Emission from SO$_2$ has also been associated with jets and outflows of protostellar sources \citep[e.g.][]{Wakelam05,Feng20}.  Figure~\ref{gradientBso2} shows the high velocity red- and blueshifted gas from SO$_2$~(8$_{2,6}$ -- 7$_{4,4}$) for VLA~1623B (a similar trend is seen for the (blended) \HCCN\ data). Both line data show a gradient aligned with the disk minor axis, which matches expectations for high velocity material associated with the jet or outflow and not a rotating disk. Due to confusion with the circumbinary disk, we cannot identify a clear gradient for VLA~1623A.  

Alternatively, the SO$_2$ emission could be tracing infalling gas rather than outflowing gas.  \citet{Sakai14} find that sulfur-bearing species can trace the centrifugal barrier, a transition layer in disks with weak shocks due to infalling gas hitting the disk. In these cases the sulfur-bearing gas is confined near the disk \citep[e.g.][]{Sakai17}. \citet{Codella24} also find a large streamer extending south from VLA~1623B in \COO\ (2-1) and several transitions of SO, which they interpret as an accretion flow. While we do not see this streamer in our SO$_2$ gas, it is detected faintly in the (blended) \HCCN\ data towards the southern part of the primary beam.  Therefore, the SO$_2$ and (blended) \HCCN\ data may be tracing a mix of both infall and outflow.  

\subsection{Circumstellar disks}\label{circumdisk}

Circumstellar disks around protostars are often well detected using rare isotopologues of CO like \COrare\ \citep[e.g.][]{Tychoniec21}. While  \COrare\ appears to trace the outflow cavity walls on larger scales (see Figure \ref{cavitywall}), we find that the high velocity gas towards the circumstellar disks have velocity gradients consistent with Keplerian rotation indicating that the compact emission indeed arises from the circumstellar disk.

\begin{figure}[h!]
\centering
\includegraphics[width=7cm,trim=0mm 8mm 0mm 0mm,clip=true]{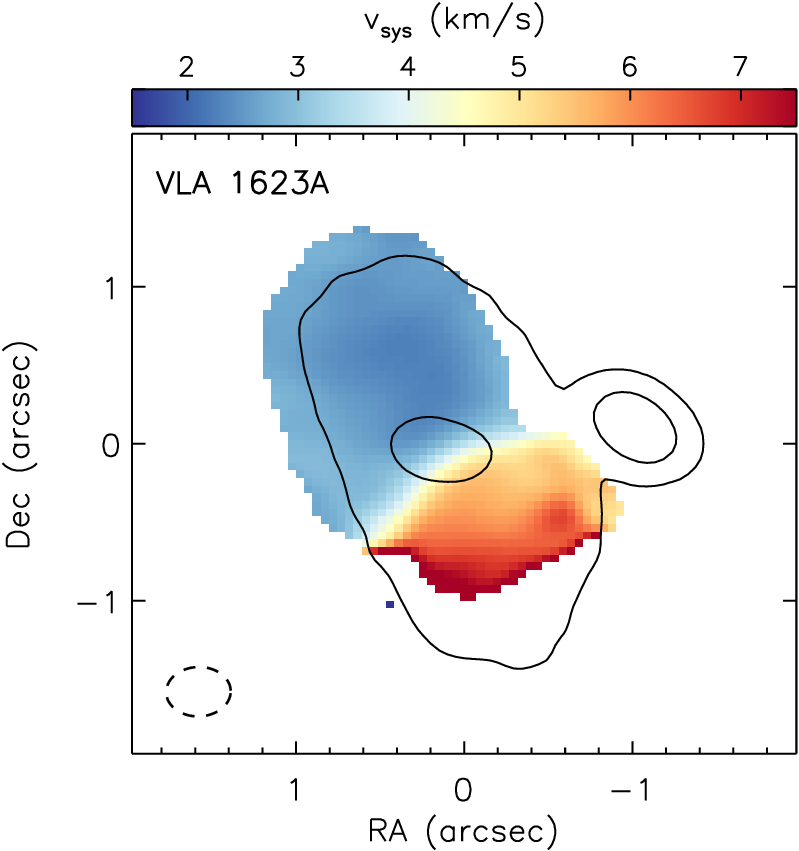}\\[3mm]
\includegraphics[width=7cm,trim=0mm 8mm 0mm 8mm,clip=true]{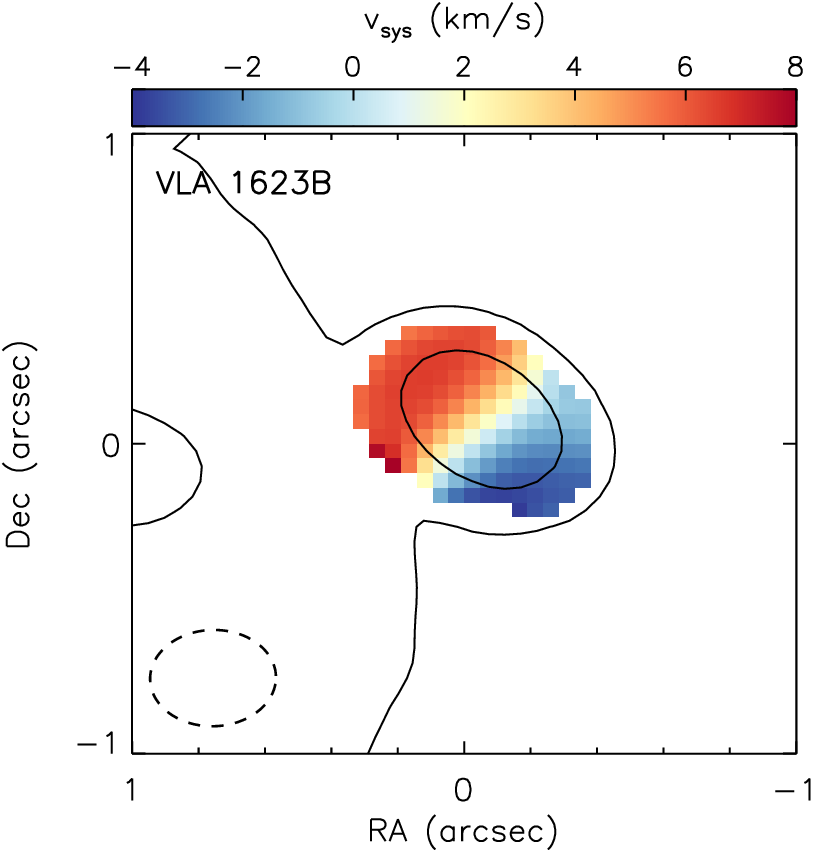}\\[3mm]
\includegraphics[width=7cm,trim=0mm 0mm 0mm 8mm,clip=true]{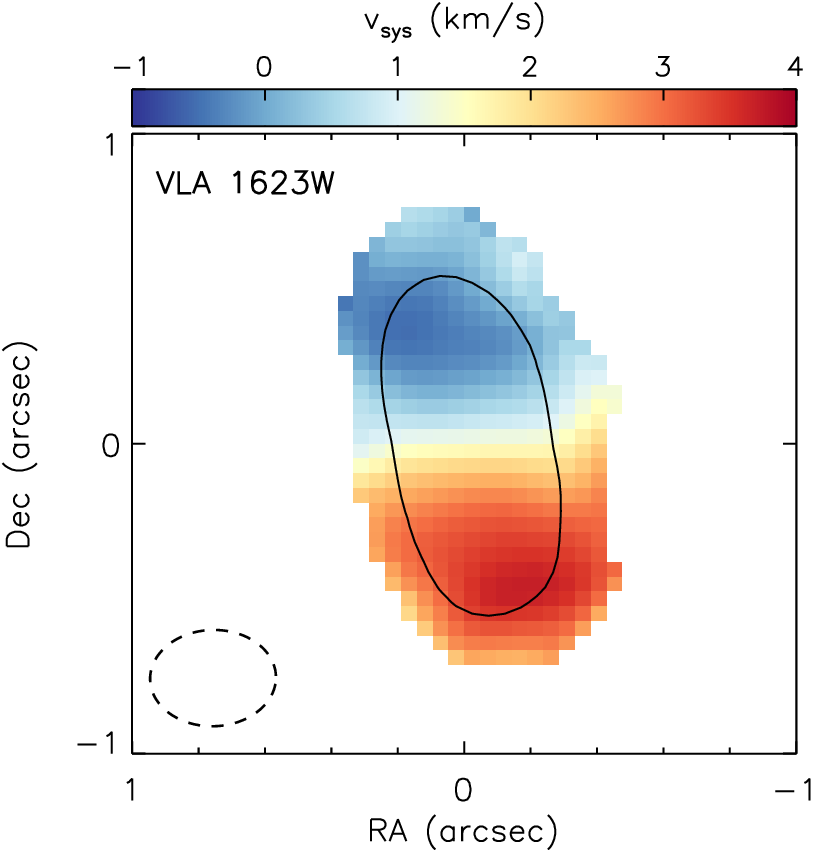}
\caption{\COrare\ Moment 1 maps for all three disks using restricted velocity channels (see the main text for details).  Black contours show the Stokes~I continuum at 7  and 49~\mJybeam.  The beam for each map is in the lower-left corner.  The Moment 1 maps are masked using thresholds in Moment 0 and dust continuum to avoid noisy pixels and cleanly show the gradients in the disks.   
\label{c17o_mom1}}
\end{figure}

Figure~\ref{c17o_mom1} shows \COrare\ (3 -- 2) Moment 1 maps of all three disks.  To ensure a good signal-to-noise, we spectrally smoothed the data to 0.4 \kms\ channels and we  calculated the Moment 1 velocities using \texttt{immoments} in CASA with strict channel selections to avoid lower velocity gas that could be confused with emission outside of the disks.  The velocity ranges used were $0.8-3.2$ \kms\ and $5.6-7.2$ \kms\ for VLA~1623A, $(-5.2)-(-2.4)$ \kms\ and $6.8-9.6$ \kms\ for VLA~1623B, and $(-2.0)-5.2$ \kms\ for VLA~1623W.  Moreover, we masked each map using thresholds in Moment 0 and dust continuum to isolate the Moment 1 data for each disk.   

All three disks show blue-red gradients indicative of rotation.  For VLA~1623A, the gradient is for the circumbinary disk as we lack the spatial resolution to resolve any gradients in the circumstellar disks of Aa and Ab.  For VLA~1623B, the \COrare\ emission is complicated by gas emanating from the circumbinary disk (see Figure \ref{mom0}).  Nevertheless, Figure \ref{c17o_mom1} shows that we can use high velocity (e.g. $> 4$ \kms\ from {the systemic velocity}) \COrare\ emission to help isolate its circumstellar material. The velocity gradient seen in VLA~1623B aligns well with the major axis of the disk. \citet{Ohashi22} find a similar gradient in VLA~1623B with high-velocity CS (5 -- 4) data. While they found that the CS data traced both the disk and outflows, the high velocity CS gas formed a gradient that was perpendicular to the outflow direction. Thus, we conclude that the high velocity \COrare\ (3 -- 2) emission is similarly tracing rotation in the disk.

Figure~\ref{c17o_mom1} also shows a clear velocity gradient for VLA~1623W going from the northern to the southern part of the disk along the major axis as expected with disk rotation.  This gradient agrees well with the \COO\ (2 -- 1) gradient presented by \citet{Mercimek23}.  We lack sensitivity in the \COrare\ data to detect a velocity gradient in the streamers like Mercimek et al., but the broad centroid velocities of the two streamers in \COrare\ (3 -- 2) match those seen in \COO\ (2 -- 1).  

In the next section, we model the \COrare\ (3 -- 2) emission using disk models with Keplerian rotation.  Broadly, we found that the emission from the highest velocity channels (greatest deviation {from the systemic velocity}) follow a $r^{-0.5}$ profile, as expected for Keplerian rotation in disks.  Nevertheless, other dynamical processes such as infall from the envelope or streamers can produce gradients. While such features can be tested using rotation curves \citep[e.g. following][]{Ohashi14,Aso15,Maret20}, the disks are sufficiently resolved or confused with circumbinary material such that the centroid position becomes unreliable.  For simplicity with our modelling, we assumed that the \COrare\ (3 -- 2) emission solely traces the rotationally supported disks and that any potential contributions from the surrounding envelope or gas streamers are negligible. To ensure this assumption is appropriate, we employed velocity cuts similar to Figure \ref{c17o_mom1} to exclude the low-velocity emission that could be dominated by these components (see Section \ref{modeling} for details).


\section{Modelling}\label{modeling}

We modelled the \COrare\ (3 -- 2) data for VLA~1623A, VLA~1623B and VLA~1623W using the \texttt{pdspy} code, which fits Keplerian rotating disk models to spectral line datasets directly in the $uv$ plane \citep{Sheehan19}. \texttt{pdspy} uses radiative transfer codes and Bayesian sampling to identify a best-fit disk model and the full posterior distributions for the parameters given a set of observations. Full details of how the code works can be found in \citet{SheehanEisner18}, but we describe the technique briefly here, including any differences from what has previously been documented.  In short, \texttt{pdspy} generates a 2D axisymmetric model of a protostellar disk. The surface density is described by a power-law function,
\begin{equation}
    \Sigma = \Sigma_0 \left(\frac{R}{R_{disk}}\right)^{-\gamma},
\end{equation}
where $\Sigma_0$ is the surface density normalization, $R$ is the radial distance from the star in cylindrical coordinates, $R_{disk}$ is the radius where the disk is truncated, and $\gamma$ is the surface density power-law exponent. For VLA~1623B and VLA~1623W, we fixed $R_{in}$, the inner radius where the disk is truncated, to 0.1 au. As VLA~1623A is a close separation binary with evidence of a cavity \citep{Harris18}, we left $R_{in}$ as a free parameter. Rather than fitting for $\Sigma_0$, we instead integrated the surface density to calculate the total (gas and dust) mass, $M_{disk}$, and used that as a free parameter in our fits.  We assumed the temperature follows a power law profile in radius, $T(R) = T_0 \left(R \, / \, 1\  \mathrm{au}\right)^{-q}$, and that the disk is vertically isothermal. The scale height of the disk as a function of radius could then be derived from the temperature at a given radius and the equations of hydrostatic equilibrium, and similarly the volume density follows from the surface density and scale height. We modelled the \COrare\ (3 -- 2) emission (no dust) assuming a constant C$^{17}$O abundance with a factor of 1/2240 relative to CO and a fixed CO abundance of $10^{-4}$ relative to molecular hydrogen \citep[e.g. following][]{Czekala15}.  We also adopted a constant turbulent velocity, $a_{turb}$, throughout the disk but leave its value as a free parameter. \texttt{pdspy} then uses the radiative transfer modelling code \texttt{RADMC-3D} and ray tracing to generate synthetic channel maps of the model disk with a given inclination ($i$) and position angle (p.a.), and then Fourier transforms the synthetic channel maps into the $uv$ plane using the \texttt{galario} code \citep{Tazzari18}. This last step ensures that the synthetic model data match the spatial scales covered by the observations. We note that the position angle is defined based on the direction of the angular momentum vector of the disk, east of north, following \citet{Czekala15}.

The model parameter values as well as the source system velocity, $v_{sys}$, and location in the field ($x_0$, $y_0$) are optimized to the observations. Unlike in \citet{Sheehan19}, we use the \texttt{dynesty} code \citep{Speagle20} instead of a Monte Carlo Markov chain (MCMC) to find the best-fit values. \texttt{dynesty} uses nested sampling to integrate the likelihood function and calculate the Bayesian evidence and the posteriors for each parameter. We opted to use \texttt{dynesty} here because we find nested sampling to be more robust with multi-modal posteriors and it also has well-defined stopping criteria.

Since the C$^{17}$O (3 -- 2) observations include emission from VLA~1623A, B, and W, we applied additional constraints on the model and data to ensure that we fit each source separately. First, we provided \texttt{pdspy} with relatively strict priors on the location of the source and allowed that position to vary minimally (by $< \pm 0.3\arcsec$ in either direction) so that the code is forced to fit the emission from the target of interest.  Second, we restricted the channels and $uv$ ranges to select emission where the target source is dominant or at least clearly separable from all other sources as well as from the envelope and outflows. For the channel restrictions, we excluded velocities between 3.25 and 5.5~\kms\ from the fit to A and between -2.5 and 6.75~\kms\ from the fit to B. For the $uv$ restriction, we excluded baselines with $uv <100$~k$\lambda$ ($\gtrsim2\arcsec$) for B and $uv <200$~k$\lambda$ ($\gtrsim1\arcsec$) for W.  The $uv$ cuts are less severe for B because there is less extended emission in the vicinity of B once the channel cuts previously described are taken into account. For VLA~1623A, we are fitting the extended circumbinary disk and as such, the velocity channel cuts are sufficient to exclude the envelope-scale without additional restrictions in the $uv$ range.  To ensure that these choices do not affect the fit quality, we carefully check the results of the models. 

Table \ref{table:best_fits} lists the best-fit model parameters for VLA~1623A, B and W. The best-fit  values are determined from the maximum likelihood models from the posterior and the uncertainties are calculated as the range around the best-fit values that include 68\% of the posterior samples.  We also added a 10\% uncertainty in quadrature to most parameters to represent additional uncertainties on those values imparted by systematic errors in, for example, flux calibration and source distance. We excluded this additional uncertainty from $x_0$, $y_0$, $i$, $p.a.$, and $v_{sys}$ as they are primarily geometric and are therefore less likely to be impacted by the aforementioned systematics. There may be additional systematic effects such as the choice of model that lead to larger errors on the measured values than those presented here \citep[e.g.][]{Premnath20}. We further note that several of the best-fit parameters are not well constrained or have unusually low or high values.  For example, some of the best-fit models have very low power-law exponents for $q$ and $\gamma$, implying flat profiles for temperature or surface density. We caution against over-interpreting the value of these disk parameters given that we had to impose strict velocity and $uv$ limits.  These cuts will limit the spatial extent of the disk, which will affect our ability to constrain the radial profiles (e.g. power-law indices). We focus instead on the results for the stellar mass, which we can reasonably constrain with the high-velocity gas, but we present the best-fit values for the other disk parameters for completeness. 

{\setlength{\extrarowheight}{5pt}%
\begin{table}
\caption{Best-fit Model Parameters\label{table:best_fits}}
\begin{tabular}{l|ccc}
\hline
\hline
Parameters & A & B & W \\
\hline
$M_{*}$ (M$_{\odot}$) & $0.27^{+0.03}_{-0.03}$ & $1.9^{+0.3}_{-0.2}$ & $0.64^{+0.06}_{-0.06}$ \\
$M_{disk}$ ($10^{-3}$ M$_{\odot}$) & $54^{+5}_{-5}$ & $>1$ & $1.7^{+0.2}_{-0.2}$ \\
$R_{in}$ (au) & $48^{+5}_{-5}$ & 0.1 & 0.1 \\
$R_{disk}$ (au) & $316^{+32}_{-32}$ & $43^{+5}_{-5}$ & $108^{+11}_{-11}$ \\
$\gamma$ & $2.0^{+0.2}_{-0.2}$ & $0.1^{+1.0}_{-0.5}$ & $0.1^{+0.1}_{-0.1}$ \\
$T_0$ (K) & $148^{+15}_{-25}$ & $34^{+56}_{-9}$ & $898^{+116}_{-173}$ \\
$q$ & $0.38^{+0.04}_{-0.05}$ & $0.06^{+0.26}_{-0.05}$ & $0.83^{+0.09}_{-0.10}$ \\
$a_{turb}$ (km s$^{-1}$) & $0.68^{+0.07}_{-0.07}$ & $0.19^{+0.24}_{-0.08}$ & $0.56^{+0.07}_{-0.07}$ \\
$i$ ($^{\circ}$) & $58.9^{+0.6}_{-0.3}$ & $91^{+16}_{-2}$ & $74.8^{+0.5}_{-0.5}$ \\
p.a. ($^{\circ}$) & $296.27^{+0.01}_{-0.12}$ & $136^{+1}_{-2}$ & $281.1^{+0.4}_{-0.5}$ \\
$v_{sys}$ (km s$^{-1}$) & $3.998^{+0.002}_{-0.002}$ & $2.31^{+0.09}_{-0.09}$ & $1.75^{+0.02}_{-0.01}$ \\
$x_0$ (mas) & $-39.4^{+1.3}_{-0.7}$ & $52^{+7}_{-6}$ & $-82^{+3}_{-3}$ \\
$y_0$ (mas) & $-90.3^{+1.1}_{-0.8}$ & $-23^{+5}_{-5}$ & $-11^{+3}_{-3}$ \\
\hline
\end{tabular}
Note: An additional 10\% uncertainty has been added in quadrature to the best-fit values for all parameters except $i$, $p.a.$, $v_{sys}$, $x_0$, and $y_0$ (see text for details).
\end{table}
}

\begin{figure*}
    \centering
    \includegraphics[width=7.2in]{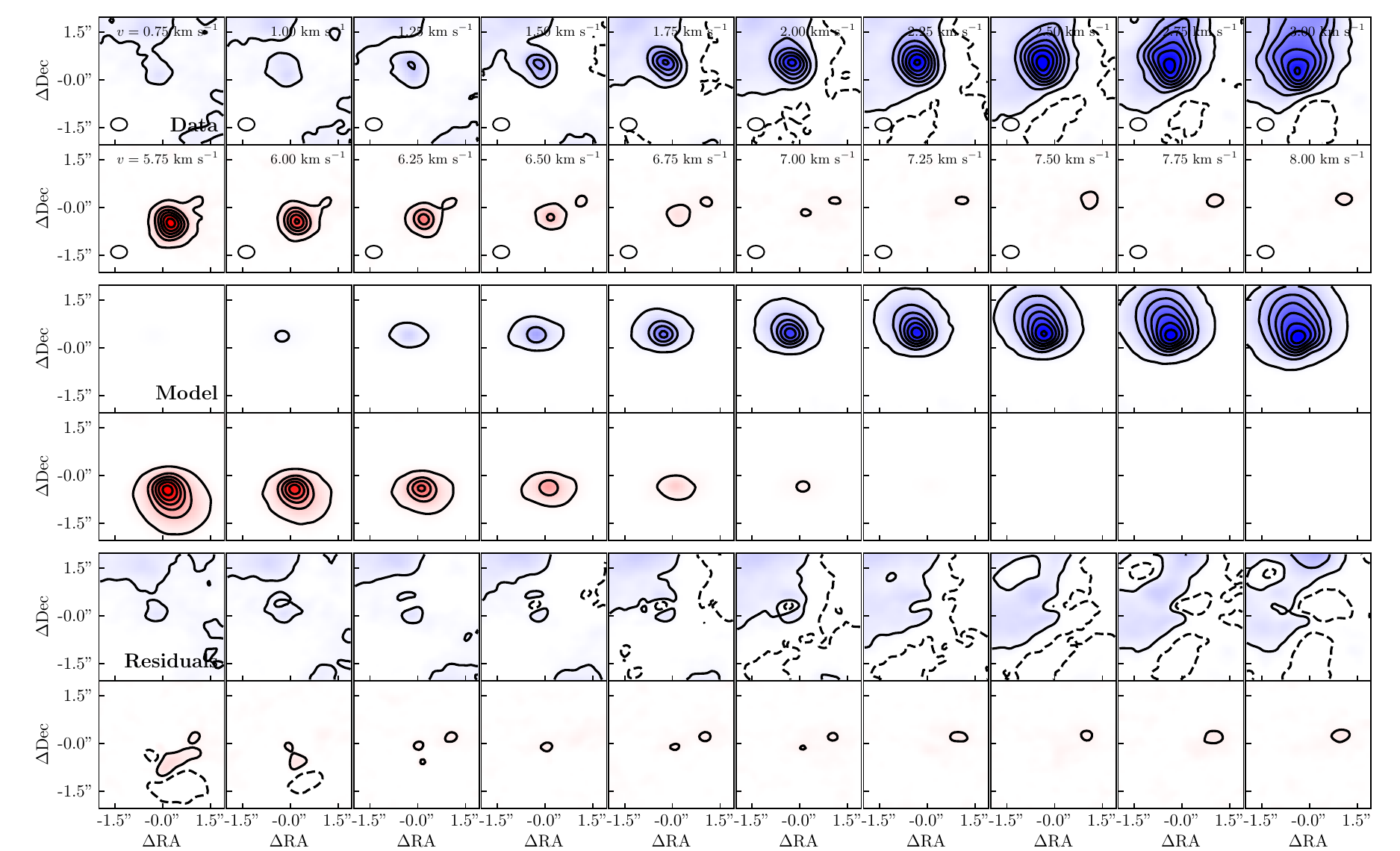}
    \caption{Comparison of the channel maps for VLA~1623A with the best-fit Keplerian disk model. The top two rows show (blue- and redshifted) channel maps from the observations,  the middle two rows show the best-fit model, and the bottom two rows show the residuals. The central velocities of the channel maps are indicated in the top two rows. For the model and residual channel maps, we used \texttt{galario} to match baselines and then  cleaned the synthetic data with similar imaging parameters as the data. The contours start at $5\sigma$ and continue in increments of $20\sigma$, with dashed contours showing negative levels. We excluded channels between 3 -- 5.75 km s$^{-1}$ as these channels were not fit for VLA~1623A.}
    \label{fig:vla1623a-best-fit-channels}
\end{figure*}

\begin{figure*}
    \centering
    \includegraphics[width=7.2in]{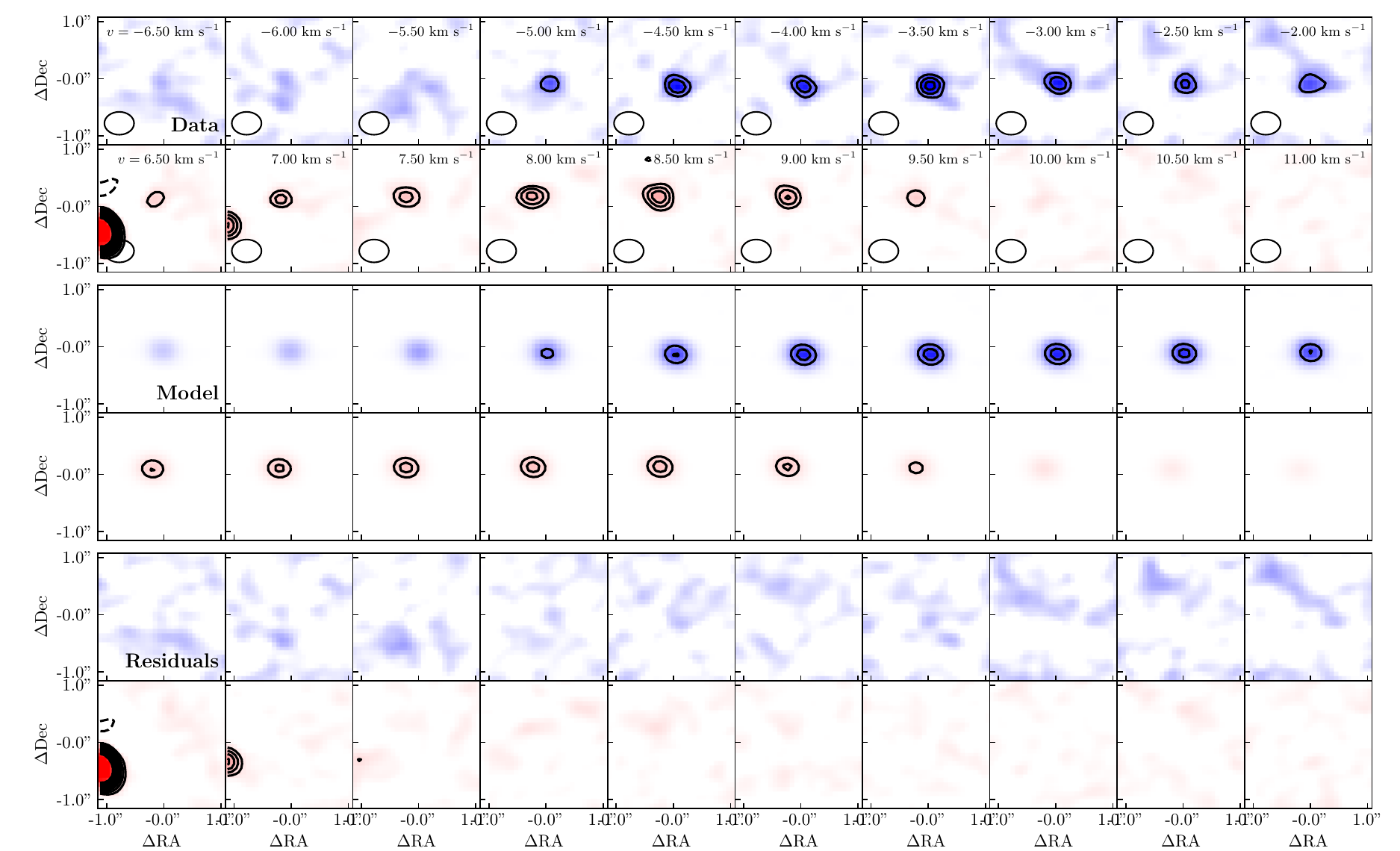}
    \caption{
    Same as Figure \ref{fig:vla1623a-best-fit-channels}, but for VLA~1623B.  Contours start at $5\sigma$ and continue in increments of $3\sigma$. Channels between -2 and 6.5 km s$^{-1}$ and baselines $<100$ k$\lambda$ are excluded. }
    \label{fig:vla1623b-best-fit-channels}
\end{figure*}

\begin{figure*}
    \centering
    \includegraphics[width=7.2in]{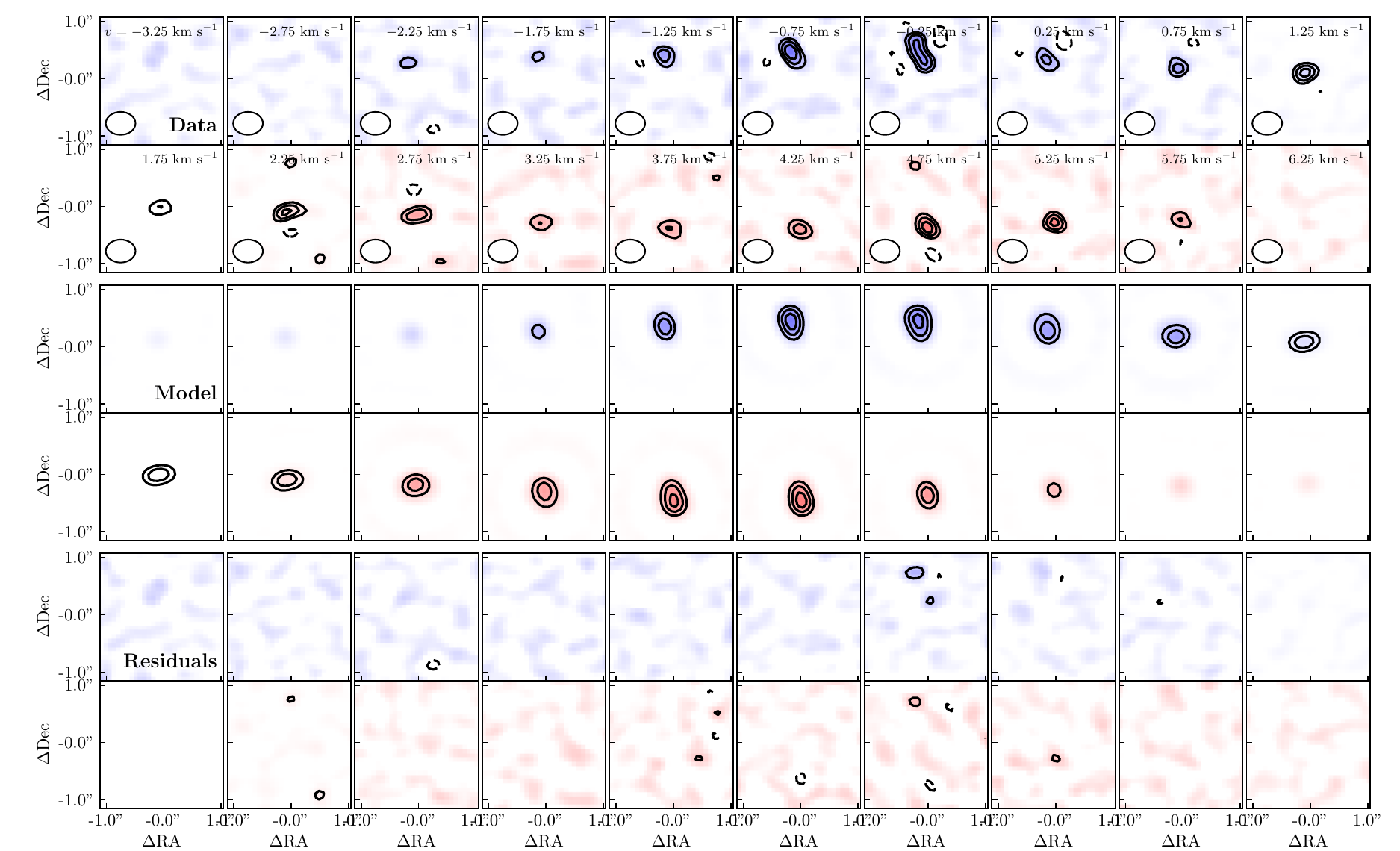}
    \caption{Same as Figure \ref{fig:vla1623a-best-fit-channels}, but for VLA~1623W.  Contours start at $5\sigma$ and continue in increments of $3\sigma$. Baselines $<200$ k$\lambda$ are excluded from the imaging, and the data are binned into $0.5$ \kms channels to improve image signal-to-noise.
    }
    \label{fig:vla1623w-best-fit-channels}
\end{figure*}

Figures \ref{fig:vla1623a-best-fit-channels}, \ref{fig:vla1623b-best-fit-channels}, and \ref{fig:vla1623w-best-fit-channels} compare the channel maps from the observations with the synthetic channel maps of the best-fit model to demonstrate the fit quality for each source. 
We find that the models fit the data well, with little or no contours above the $5\sigma$ level appearing in the residual channel maps associated with the target disk. Emission that exceeds $5\sigma$ in the residuals generally arises from structures not included in the model but are present in the field of view.  For example, Figure \ref{fig:vla1623b-best-fit-channels} shows significant redshifted residual emission above the 5$\sigma$ level in the channel maps of VLA~1623B, but this residual emission comes from the circumbinary disk, which is not included in the model of VLA~1623B. This broad agreement between the observations and the model indicates that our simple model of Keplerian rotation was indeed reasonable and that other dynamical processes such as infall are either negligible or limited to larger scales than that of the disk.

\begin{figure*}
    \centering
    \includegraphics[width=2.38in]{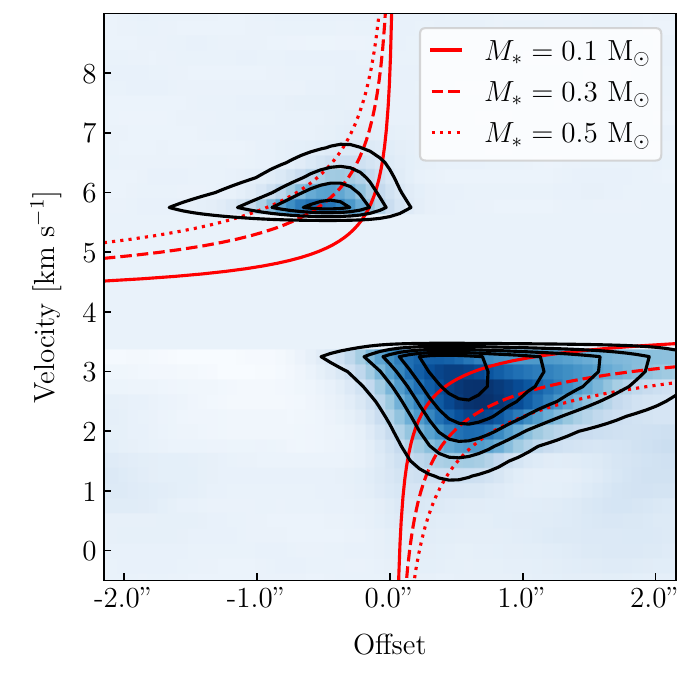}
    \includegraphics[width=2.38in]{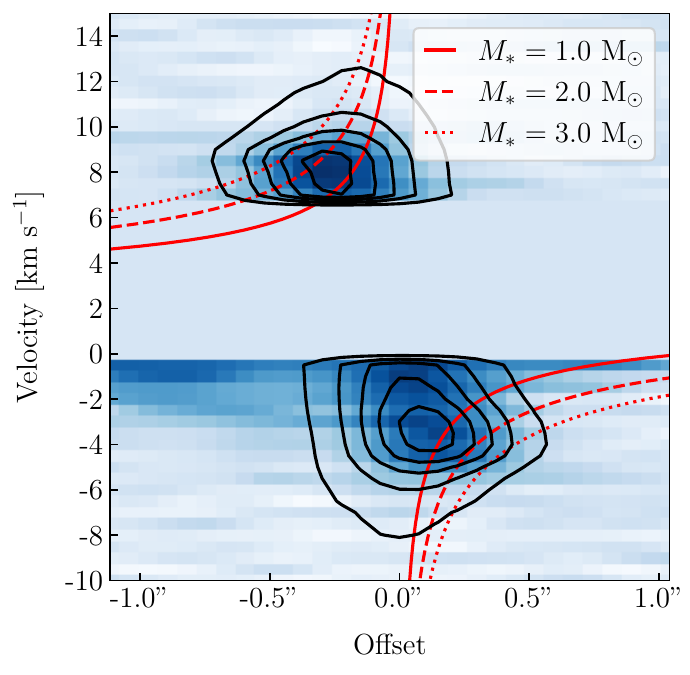}
    \includegraphics[width=2.38in]{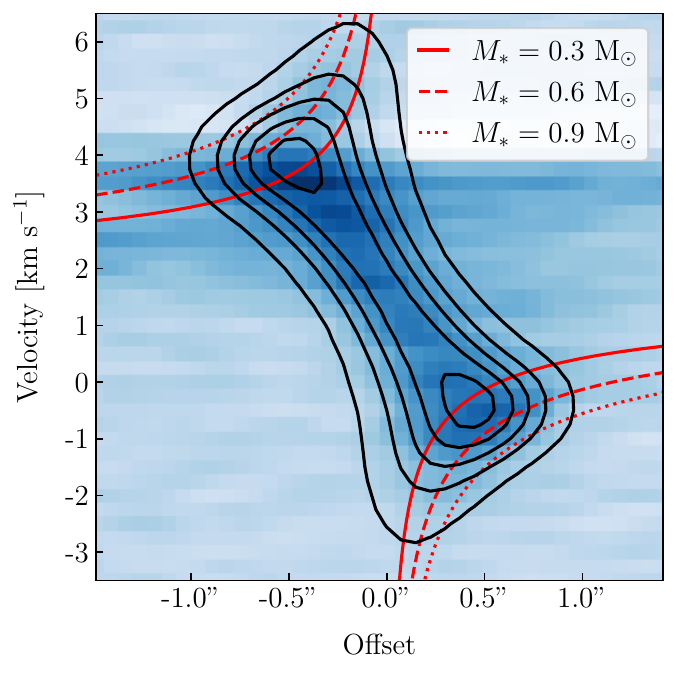}
    \caption{Position-velocity (PV) diagrams for VLA~1623A ({\it left}), VLA~1623B ({\it centre}), and VLA~1623W ({\it right}).  The background images show the observed data, and  contours show the synthetic data from the best-fit model for each source.   The PV diagrams were extracted  with an aperture width of $\sim0.35\arcsec$, i.e. about one beam-width, perpendicular to the extracted spatial direction for VLA~1623B and VLA~1623W as the emission for both sources is marginally resolved. As VLA~1623A is much better resolved, we instead used a width of $\sim0.7\arcsec$ to extract its PV diagram. Velocities between 3.25 \kms\ and 5.5 \kms\ were masked for VLA~1623A to avoid channels with significant envelope emission, and velocities between -2.5 \kms\ and 6.75 \kms\ were masked for VLA~1623B to avoid confusion from emission associated with VLA~1623A. We note that these cuts lead to asymmetries in the number of channels available on either side of the line centre.  We also show example Keplerian rotation curves for stars with a range of stellar masses as a simple visual comparison.}
    \label{fig:pvdiagram-best-fits}
\end{figure*}

Figure \ref{fig:pvdiagram-best-fits} compares the observed  position-velocity data along the disk major axis (background colour) with the synthetic position-velocity data (contours) from the best-fit models of VLA~1623A, B, and W. Though position-velocity diagrams represent a reduction of the dimensionality of the data, they can be a useful tool for by-eye comparison with Keplerian rotation.  Figure \ref{fig:pvdiagram-best-fits} shows simple Keplerian rotation curves for fixed stellar masses to demonstrate that the masses derived from our fitting are consistent with what might be obtained from such a simple comparison.

\section{Discussion} \label{discussion}

\subsection{Stellar masses}

Previously, Keplerian rotation modelling had only been done for VLA~1623A.  \citet{Murillo13} determine a (combined) stellar mass of $M_\star \sim 0.2$ \Msun\ using models of Keplerian rotation with infall and early (Cycle 0) ALMA \COO\ (2 -- 1) data.  With more sensitive data, \citet{Hsieh20} measure a combined stellar mass of $M_\star \sim 0.3-0.5$ \Msun\ for VLA~1623A.  In both cases, the mass values for VLA~1623A are a total mass for the tight binary, VLA~1623Aa and VLA~1623Ab \citep{Harris18}.  The remaining stellar components, VLA~1623W and VLA~1623B, have estimated masses from the velocity gradient assuming Keplerian rotation, rather than fitting the Keplerian motion itself.   For VLA~1623B, \citet{Ohashi22} measure a mass of $M_{\star} \lesssim 1.7$ \Msun\ from CS (5 -- 4) data with $v_{rot} = 7.8$ \kms\ at $R = 27$ au, and for VLA~1623W, \citet{Mercimek23} estimate a mass of $M_{\star} \approx 0.45$ \Msun\ from \COO\ (2 -- 1) data with $v_{rot} \approx 3$ \kms\ at $R \approx 50$ au.    

Broadly, our measured stellar masses from the radiative transfer models agree well with the previously inferred dynamical masses.  We find best-fit stellar masses of $M_\star = 0.64 \pm 0.06$ \Msun\ for VLA~1623W  and $M_\star = 1.9^{+0.3}_{-0.2}$ \Msun\ for VLA~1623B, and a combined stellar mass of $M_\star = 0.27 \pm 0.03$ for VLA~1623A (see Table \ref{table:best_fits}).  Since these masses are measured from \COrare\ (3 -- 2), we have a rarer isotopologue of carbon monoxide than \COO\ and less confusion with the infalling envelope and streamers \citep{Murillo13,Mercimek23}.  Moreover, our modelling process fits the full 3D (position, position, velocity) dataset with physically motivated models rather than estimating the mass from  the peak line emission in a few channels at a radius offset from the disk centre. In addition, \texttt{pdspy} fits visibility data.  The visibility plane encodes information at several times smaller spatial scales than is recovered by deconvolution \citep{Jennings20}, and as such we are not limited by the beam resolution and can better constrain the models on sub-beam scales.  

These measurements put constraints on the stellar masses for all the components (with VLA~1623A combined) and indicate a hierarchy.  VLA~1623B has the highest mass, with more than twice the mass of the other two stellar components combined.  VLA~1623W is second in mass with VLA~1623A in third.  These more accurate stellar masses yield valuable insights into the physical conditions of VLA~1623, and its formation and evolution.   We discuss individual scenarios in Section \ref{sec:origin}.

\subsection{Origin of the VLA 1623 system}\label{sec:origin}

The detection of Keplerian rotation towards both VLA~1623W and VLA~1623B indicates that these sources are genuine protostars \citep[see also][]{Ohashi22, Mercimek23}.  Nevertheless, the VLA~1623 system is a complex environment and its origin is not straightforward.  Here, we consider different scenarios to explain the formation and evolution of VLA~1623.

\subsubsection{Dynamical interactions}\label{dis:eject}

One theory to explain the VLA~1623 system is that VLA~1623W was ejected.  An ejection scenario has been proposed because the proper motions of VLA~1623W indicate that it points back to the VLA~1623A/B protostars \citep{Harris18}.  A dynamical interaction could also explain the counter-rotating disks between VLA~1623A and VLA~1623B \citep{Takaishi21, Ohashi22}, the lack of envelope emission around VLA~1623W, and the large gas streamers \citep{Murillo13, Mercimek23}.

Nevertheless, there are challenges to explaining VLA~1623W as an ejected object.  First, multi-body interactions generally result in the ejection of the lowest-mass object \citep{Reipurth10}.  From our stellar mass measurements, VLA~1623A has the lowest mass by more than a factor of two even if the circumbinary disk is included. Second, a dynamical interaction leading to ejection is expected to produce a spiral structure in the circumbinary disk \citep{Takaishi21}, which is not observed, although a spiral structure caused by dynamical interactions could dissipate quickly \citep{Cuello22}. Third, VLA~1623W does not appear to get close enough to VLA~1623A/B for a dynamical ejection to be likely.  In Appendix \ref{app_proper}, we extrapolate the proper motion of VLA~1623W and VLA~1623B back in time to find a closest plane-of-sky encounter of about 650 au. \citet{Reipurth10} simulated dynamical interactions of triple systems of protostars in dense cores at separations between 50 au and 400 au and found that ejections were weaker and less common at 400 au than at 50 au \citep[see also the results from stellar flybys from][]{Cuello22}.  Our analysis shows that the closest separation only becomes $< 100$ au at the 6th percentile, whereas the closest separation is $\gtrsim 400$ au at the 25th percentile (see Appendix \ref{app_proper}). For such wide projected separations, we would expect that the stars would be too distant to have a substantial gravitational effect on their relative orbits \citep{SadavoyStahler17}.  As a result, a dynamical ejection scenario appears unlikely to explain the VLA~1623 system.  

\subsubsection{Chance alignment}\label{dis:chance}

Since VLA~1623W has a different system velocity relative to VLA~1623A and B, it may be an unrelated young stellar object (YSO) that happens to have a chance alignment with the VLA~1623A/B system.  \citet{MurilloLai13} identify VLA~1623 as having non-coeval YSO components based on their SEDs \citep[see also][]{Murillo18}.  VLA~1623W also lacks envelope emission and an outflow \citep{Murillo13, Santangelo15,Hara21}, which could indicate that it is a more evolved YSO that formed separately.

Following \citet{Tobin22}, we estimated the probability that VLA~1623W is a companion source using Bayes theorem,  
\begin{equation}
    P(c|d) = \frac{P(d|c)P(c)}{P(d)}
,\end{equation}
where $P(d|c)$ is the probability we would detect a true companion source, $P(c)$ is the probability of having a companion at a given separation, and $P(d)$ is the probability of a detection of any YSO.  We adopted $P(d|c) = 0.75$, similar to \citet{Tobin22}.  \citet{Tobin22} measured their companion detection probability based on a dust mass sensitivity of $\sim 1$ M$_\oplus$ for dusty disks. Since we have higher sensitivity observations of $\sim 0.01$ M$_\oplus$ for source detection \citep[e.g.][]{Sadavoy19}, our adopted value of $P(d|c)$  should be considered a lower limit.  We also used $P(c) = 0.2$ from \citet{Tobin22}, which is based on companion statistics from Perseus and Orion for a separations $\lesssim 1000$ au.  Finally, we calculated the probability of having a detection, $P(d)$, with
\begin{equation}
    P(d) = 0.75P(c) + \left[1 - e^{-0.75\Sigma\pi r^2}\right]\left[1 - 0.75P(c)\right]\label{eq.Pd}
,\end{equation}
where $\Sigma$ is the stellar surface density and $r$ is the separation being considered. Equation \ref{eq.Pd} gives the probability of detecting any source at the observed sensitivity.  The first term is the likelihood of detecting a true companion, assuming 75\%\ are detectable at our sensitivity, and the second term is the likelihood of detecting an unrelated source in a specified area for a given YSO stellar density \cite[for further details, see Appendix A of][]{Tobin22}.  We used $r = 10\arcsec$ for the area. To estimate the YSO stellar density, we calculated the 11th nearest neighbour at the position of VLA~1623 from the \emph{Gaia}-corrected YSO catalogue of \citet{Grasser21}.  The 11th nearest neighbour gives $\Sigma_{11} = 1470$ YSO pc$^{-2}$. This stellar density is slightly below the peak YSO stellar density for L1688 of 2000 YSOs pc$^{-2}$ from \citet{Gutermuth09}, but VLA~1623 is located off the cluster centre and so the 11th nearest neighbour should be more representative. 

With our assumed values, we calculate a probability that VLA~1623W is a companion protostar of VLA~1623 of $P(c|d) = 0.55$, which suggests that VLA~1623W has nearly equal chances to being a member or an unrelated YSO.  Nevertheless, this probability of 55\%\ is likely a lower limit.  First, VLA~1623W appears protostellar in nature (see Section \ref{dis:insitu}), and the stellar surface densities from both \citet{Gutermuth09} and \citet{Grasser21} are dominated by Class II sources.  Since protostars (embedded YSOs) are less common than Class II YSOs \citep[e.g.][]{Gutermuth09, Dunham15}, the probability that VLA~1623W is a companion object should be higher than our current estimate.  Second, the streamers towards VLA~1623W indicate that it is interacting with the dense core itself and therefore is likely within the dense core and not completely unrelated to the VLA~1623 system. These factors increase the likelihood that VLA~1623W is a true companion source, so we consider VLA~1623W to be a member YSO.

\subsubsection{In situ formation}\label{dis:insitu}

Here, we define objects that formed out of the same natal core environment as forming in situ.  In general, there are two main mechanisms that have been proposed to produce multiple stellar systems: disk fragmentation and turbulent fragmentation \citep{Offner23}. Briefly, disk fragmentation occurs when a circumstellar disk becomes unstable to gravity and fragments to form additional stars, primarily at small ($\lesssim 100$ au) separations \citep[e.g.][]{Bonnell94, Kratter10}.   Turbulent fragmentation occurs when density perturbations within the natal dense core become massive enough for self-gravity, allowing them to collapse and form independent objects at a wide range of separations \citep[e.g.][]{Offner10, Offner16, Kuffmeier19}.  Although the separation distributions can broadly indicate the formation mechanism, simulations show that wide binary systems can shrink to smaller orbits on timescales of $\lesssim 0.1$ Myr \citep{Offner12}, which is less than the protostellar lifetime.  As a result, differences in angular momentum vectors may be more illuminating, as disk fragmentation predicts aligned vectors and turbulent fragmentation implies random vectors \citep[see][and references therein]{Offner23}.  

VLA~1623A and VLA~1623B have a projected separation of roughly 200 au, whereas VLA~1623W is over 1000 au away from the pair.  The wide separation for VLA~1623W would indicate that it formed via turbulent fragmentation, but from separation alone, it is difficult to conclude the origins of A or B, especially given the size of the circumbinary disk.  Nevertheless, VLA~1623A and VLA~1623B have misaligned velocity gradients indicative of counter-rotation \citep[see Figure \ref{c17o_mom1} and][]{Ohashi22} and the disk inclinations and positions angles do not agree as well (see Table \ref{table:best_fits}).  Both of these factors are consistent with turbulent fragmentation.  Moreover, since VLA~1623B is more massive than VLA~1623A and its circumbinary disk, it seems unlikely that it was formed via disk fragmentation and then was later perturbed through dynamical interactions to have a misaligned rotation axis.  

Turbulent fragmentation does not require that all stellar components formed at the same time. \citet{MurilloLai13} and \citet{Murillo18} compared the SEDs and circumstellar material for the VLA~1623 components and concluded that VLA~1623W appeared more evolved than VLA~1623A and VLA~1623B.  Indeed, there is a strong, collimated outflow coming from the VLA~1623A/B system, whereas VLA~1623W has no detected outflows or envelope  \citep{Murillo13, Santangelo15, Hara21, Michel22}, in agreement with this protostar being older.  Thus, VLA~1623W may have formed first through turbulent fragmentation and then VLA~1623A/B formed later.

Although the protostars in VLA~1623 may not be entirely coeval, this does not mean that VLA~1623W must be an evolved YSO (e.g. Flat or pre-main sequence).   First, VLA~1623W is detected in \COO\ \citep{Mercimek23} and \COrare.  Protostellar disks are generally warmer, which favour the detection of rare isotopologues in the gas phase, and these disks also tend to be more massive and optically thick, which will shield the molecules from being selectively photo-dissociated as seen in older protoplanetary disks \citep[e.g.][]{Miotello17,vantHoff18,Zhang20, ArturV19}.  Some protoplanetary disks have been detected in \COrare\ \citep[e.g.][]{Zhang21}, but these observations are generally towards the inner radii, whereas we detect \COrare\ towards the entire disk of VLA~1623W (e.g. beyond the dust disk; see Figure \ref{c17o_mom1}).  Younger protostars tend to be brighter with warmer disks, which means that rare CO isotopologues will be detected in the gas phase out to larger radii \citep[e.g.][]{vantHoff18}. Second, VLA~1623W appears to be flared at 0.87~mm and 1.3~mm \citep{Michel22}, indicating that larger dust grains have not yet had time to settle to the midplane.  Protoplanetary disks tend to be geometrically thin at those wavelengths \citep[e.g.][]{Villenave20}, whereas flaring has been detected towards some protostellar disks \citep[e.g.][]{Sheehan22_l1527,Ohashi23}.  Finally, VLA~1623W is an optically thick, relatively massive disk \citep{Harris18,Sadavoy19,Michel22}, and disk mass tends to decrease with evolutionary stage for a given star-forming region \citep[e.g.][]{Tobin20}.  Table \ref{table:best_fits} gives an equivalent disk dust mass of 6 M$_{\oplus}$, assuming a gas-to-dust mass ratio of 100.  This dust mass places VLA~1623W in the upper half of disk masses from the Ophiuchus disk survey, ODISEA, and is consistent with Class I (or Flat) disks \citep{Williams19}.  Although there are more massive disks around evolved YSOs in other clouds, the ODISEA survey shows that disk masses in Ophiuchus tend to be lower than other nearby systems, making the mass of VLA~1623W significant relative to the other Ophiuchus disks.

Even though it has a youthful disk, VLA~1623W lacks an envelope component typical of protostellar sources. Since VLA~1623W is located along the collimated outflow axis (in projection), its envelope could have been stripped \citep[e.g.][]{deGouveia05,Ladd11}.  In this case, we would observe a less embedded SED and the protostar would have less accretion to drive an outflow.  Currently, there is no evidence of shocked gas towards VLA~1623W (e.g. it is not detected in SO$_2$), although we could lack the sensitivity.   While we cannot clearly classify VLA~1623W given the uncertainties from its SED and challenges with its high disk inclination, we believe that VLA~1623W is more likely to be a protostellar source and less likely to be a more evolved (e.g. Class II) object.  

For the origin of the VLA~1623 protostellar system, we propose that the main stellar components, VLA~1623A, VLA~1623B, and VLA~1623W, formed via turbulent fragmentation, whereas the tight VLA~1623A binary system (Aa and Ab) at the centre of a large circumbinary disk may have formed via disk fragmentation given their small separation, although we cannot rule out that all the stellar components formed via core fragmentation and migrated.  A full origin picture must account for the wide range of stellar masses for each of the stars, explain why the circumbinary disk formed around the lowest-mass stellar companion, and determine the stability of the stellar system and the circumbinary disk.

\subsection{VLA~1623 stability}

In this section, we consider the gravitationally stability of the protostellar disks and the VLA~1623 system.  For disk stability, we used the disk-to-star mass ratio to assess the gravitational stability of the circumstellar and circumbinary disks. Systems with low disk-to-star mass ratios should be more stable to fragmentation, whereas higher ratios would be unstable \citep[e.g.][]{Vorobyov10}.   \citet{MercerStamatellos20} find that typical ratios of $\gtrsim 30\%$ favour fragmentation in disks based on a sample of low-mass stars.   This cutoff matches what is seen in the L1448~IRS3 system, where the fragmenting circumbinary disk of L1448~IRS3B has a large disk-to-star mass ratio of 25\%\ and the gravitationally stable disk around  L1448~IRS3A has a smaller disk-to-star mass ratio of 3\%  \citep{Reynolds21}.  Gravitational stability in L1448~IRS3 was independently assessed using a Toomre Q analysis and the presence or absence of spiral structures and fragments in the disks themselves \citep{Tobin16l1448,Reynolds21}.

From our best-fit models in Table \ref{table:best_fits}, we calculate ratios of $M_{disk}/M_\star \approx 20$\%\ for the circumbinary disk around VLA~1623A and $M_{disk}/M_\star \approx 0.3$\% for both VLA~1623B and VLA~1623W.  These results imply that the stellar disks are likely stable under their own self-gravity, even with the uncertainties on the disk masses from the models.  The circumbinary disk, however, has an intermediate ratio, which could indicate that it is close to being unstable and may undergo fragmentation in the future.    Nevertheless, there is no evidence of ongoing spiral structure or fragmentation in the circumbinary disk \citep[e.g.][]{Sadavoy18b,Harris18}, both of which are considered signposts of gravitational instabilities \citep[e.g.][]{Kratter10} and  are detected in L1448~IRS3B \citep{Tobin16l1448}.  We cannot rule out that the circumbinary disk fragmented in the past (e.g. to form the tight VLA~1623Aa and VLA~1623Ab binary) and has since reached a more stable point.  

In the case of the circumbinary disk, spiral structure could also arise due to dynamical interactions with its close companion, VLA~1623B, rather than gravitational instabilities. It is interesting that we see no evidence of perturbations in the circumbinary disk given the close (in projection) separation between VLA~1623A and VLA~1623B.  Specifically, VLA~1623B is the most massive YSO in the system (e.g. its stellar mass exceeds the estimated combined mass of VLA~1623A and the circumbinary disk by a factor of six) and the projected separation of VLA~1623A and VLA~1623B is approximately equal to the minimum value given their disk sizes.  The circumbinary disk, however, is well fit by an isolated disk model with no significant residuals indicating deviations from Keplerian rotation.  Moreover, both disks have low gas temperatures \citep{Murillo18gas}, which suggest no dynamical interactions or perturbations are taking place.  Both factors imply that VLA~1623B may have a large line-of-sight separation and is not physically close enough to VLA~1623A to perturb or heat the gas in the circumbinary disk.  A large line-of-sight separation could also indicate that VLA~1623B is an unrelated YSO, but given their close separation in projection, they are more likely to be true companion. Using the same probability definitions as Section \ref{dis:chance}, we find $P(c|d) = 0.95$ for VLA~1623A and VLA~1623B to be companion YSOs, assuming $P(c) = 0.14$ for separations $<500$ au \citep{Tobin22} and a distance of $r = 1.5\arcsec$.   With such a high probability, VLA~1623B should be considered a companion protostar.

Finally, we considered whether VLA~1623W is itself gravitationally bound to the VLA~1623A/B YSOs, assuming it is a true companion source (see Section \ref{dis:chance}).  For simplicity, we ignored any contributions from the surrounding dense core, and calculated the potential energy between VLA~1623W and VLA~1623A/B using the star and disk masses in Table \ref{table:best_fits} and a projected separation of $r=10$\arcsec, and we calculated a kinetic energy based on the relative 3D motions of the A/B and W from their system velocities (assuming 3 \kms\ for A/B and 1.75 \kms\ for W) and proper motions (see also Appendix \ref{app_proper}).  The resulting energies are $\Omega = -GM_{AB}M_W/r \approx -0.2 \times 10^{37}$ J for the gravitational energy and $K = \frac{1}{2}M_W\sigma_{3D}^2 \approx 1 \times 10^{37}$ J for the kinetic energy. These back-of-the-envelope calculations imply that the turbulent kinetic energy is roughly a factor of five larger than the gravitational energy from the YSOs alone.  Moreover, the calculated gravitational energy is an upper limit, since we only have a projected separation between VLA~1623A/B and VLA~1623W, and the true 3D distance could be much larger. Thus, the kinetic energy appears several times larger than the gravitational energy, indicating that VLA~1623W is unlikely to be bound to the VLA~1623A/B system and could disperse, although the source may still be bound to the dense core.


\section{Conclusions}\label{summary}

We present new ALMA molecular line observations of the VLA~1623 system.  We primarily focused on \COrare\ (3 -- 2) observations, which trace the disks of the protostars and show velocity gradients consistent with Keplerian rotation.  We used the radiative transfer modelling code \texttt{pdspy} to model the \COrare\ (3 -- 2) emission for VLA~1623A, VLA~1623B, and VLA~1623W, obtaining constraints for their stellar masses.  Our main conclusions are:

\begin{enumerate}

\item We measure stellar masses of 0.27 \Msun, 1.9 \Msun, and 0.64 \Msun\ for VLA~1623A (Aa and Ab combined), VLA~1623B, and VLA~1623W, respectively.  These masses are in good agreement with previous estimates that used different tracers and techniques.  

\item Based on the new mass measurements and an analysis of the proper motion of the stars, we disfavour a scenario where VLA~1623W was ejected from the central core.  

\item Following \citet{Tobin22}, we estimate a probability of VLA~1623W being a companion source of nearly 55\%, which means we cannot rule out that it is an unrelated YSO along the line of sight.  Nevertheless, based on the apparent youth of the disk and the presence of gas streamers connecting VLA~1623W to the VLA~1623A/B protostars, we favour it being a genuine companion source.   

\item We propose a scenario where VLA~1623A, VLA~1623B, and VLA~1623W formed initially from turbulent fragmentation, although VLA~1623A may have undergone disk fragmentation to produce the tight binary, Aa and Ab.  The protostars may not be entirely coeval with each other, but differences in age should not exceed the protostellar lifetime.

\item The disks around VLA~1623B and VLA~1623W appear gravitationally stable based on very low disk-to-star mass fractions. The circumbinary disk has an intermediate fraction of 20\%, which could indicate instability and future fragmentation.  Nevertheless, we see no evidence of spiral structure in the circumbinary disk, either from gravitational instability or dynamical interactions.   

\end{enumerate}

We also find that VLA~1623W appears to be unbound to the other protostars, suggesting that it may disperse. As such, these observations represent a rare snapshot in time of a multiple-protostellar system prior to its diffusion or strong dynamical interactions.  While the presence of streamers towards VLA~1623W \citep{Mercimek23} indicates that some interactions between the stars and envelope may be occurring, we do not see disturbed gas motions in the envelope or circumbinary disk. Moreover, given the mass of VLA~1623B and its projected separation, we would expect it to influence the circumbinary disk, although we detect no evidence of perturbations with the current data.  Future analyses that model the circumbinary disk may yet discover deviations from Keplerian rotation due to gravitational perturbations from the more massive VLA~1623B, which would show the onset of dynamical interactions in action.

\vspace{1cm}
\begin{acknowledgements}
We thank the anonymous referee for valuable comments that improved the publication.  The authors thank the NAASC for support with the ALMA observations and data processing.   The authors would also like to thank Ewine van Dishoeck for valuable discussions. SIS acknowledges the support for this work provided by the Natural Science and Engineering Research Council of Canada (NSERC), RGPIN-2020-03981, RGPIN-2020-03982, and RGPIN-2020-03983.  This paper makes use of the following ALMA data: ADS/JAO.ALMA\#2018.1.01089.S. ALMA is a partnership of ESO (representing its member states), NSF (USA) and NINS (Japan), together with NRC (Canada), MOST and ASIAA (Taiwan), and KASI (Republic of Korea), in cooperation with the Republic of Chile. The Joint ALMA Observatory is operated by ESO, AUI/NRAO and NAOJ.  The National Radio Astronomy Observatory is a facility of the National Science Foundation operated under cooperative agreement by Associated Universities, Inc.  This research has made use of the SIMBAD database, operated at CDS, Strasbourg, France.

\end{acknowledgements}

\bibliographystyle{aa}
\bibliography{references}

\begin{appendix}

\section{Proper motion and closest approach}\label{app_proper}

This appendix outlines the calculations for the proper motions for VLA~1623B and VLA~1623W.  We excluded VLA~1623A from these calculations since the combination of it being a tight binary and having a circumbinary disk can confuse its centroid position. We took positions of VLA~1623B and VLA~1623W from the literature that present high-resolution data ($\lesssim 1\arcsec$) and where the epoch of observations could be  identified.  Table \ref{tab_pm} lists the corresponding archival data, where Column 1 gives the reference, Column 2 gives the epoch of observation, Column 3 gives the observing frequency, Columns 4 and 5 give the source right ascension and its adopted error, Columns 6 and 7 give the source declination and its adopted error, Column 8 gives the geometric mean size of the synthesized beam, and Column 9 gives the phase calibrator for the observations.  We follow \citet{Sadavoy18b} in using the map resolution scaled by the source peak S/N to estimate the position uncertainties.  For the present study and those of \citet{Sadavoy19} and \citet{Harris18}, both sources had peak S/N $>100$, so we adopted a minimum error of 5 mas. %

{\setlength{\extrarowheight}{0.8pt}%
\begin{table*}
\caption{Literature positions of VLA 1623W.}\label{tab_pm}
\begin{tabular}{lcclclccl}
\hline\hline
Reference$^{a}$ &       Epoch   & Freq       & $\alpha$ & $\sigma_{\alpha}$      &$\delta$ & $\sigma_{\delta}$   & FWHM$^{b}$ & Phase Calibrator\\
                                &                       & (GHz)  & (h:m:s)              &  (mas)  & (d:am:as)             & (mas)         & (arcsec)              & \\
\hline
This study              & 2019.3                & 342 & 16:26:25.6295   & 5               & -24:24:29.6258&       5               & 0.27          & J1650-2943\\
Sadavoy19 (1)           & 2017.5                & 233 &  16:26:25.6315   & 5     & -24:24:29.6185 &      5               & 0.24          & J1625-2527 \\
Harris18 (2)            & 2016.3                & 344 & 16:26:25.6316    & 5     & -24:24:29.5866&       5               & 0.18          & J1625-2527, J1633-2557 \\
Murillo13$^{c}$ (3)     & 2012.33               & 230 & 16:26:25.636     & 16            & -24:24:29.488&        16              & 0.65          & J1733-1304 \\ 
Maury12$^{c}$   (4)     & 2009.58               & 225 & 16:26:25.63              & 100   & -24:24:29.5   &       100             & 0.53          & J1625-2527, J1517-2422 \\  
Chen13$^{c}$    (5)     & 2007.5                & 230 & 16:26:25.64              & 200   & -24:24:29.3   &       200             & 0.88          & J1626-2951 \\ 
\hline
\end{tabular}
\begin{tablenotes}[normal,flushleft]
\item $^a$References for positions are: (1) \citet{Sadavoy19}, (2) \citet{Harris18}, (3) \citet{Murillo13}, (4) \citet{Maury12}, (5) \citet{Chen13}.  
\item $^b$ The geometric mean FWHM ($= \sqrt{ab}$).
\item $^c$ Position offsets are estimated using total flux
\end{tablenotes}
\end{table*}
}

\begin{figure}[h!]
\includegraphics[width=0.485\textwidth]{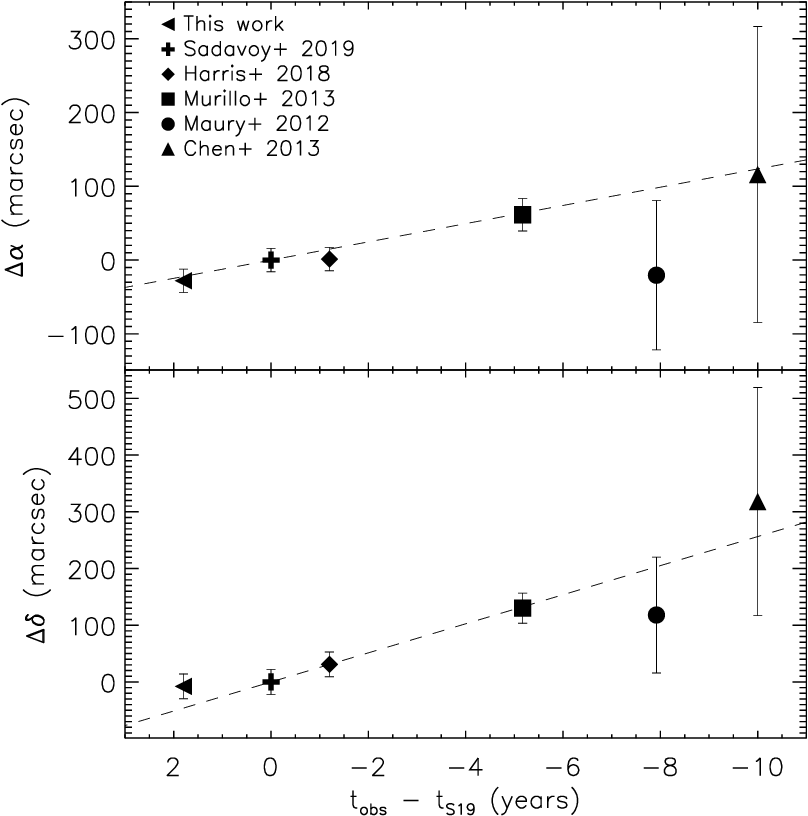}
\caption{Proper motion fits for VLA 1623W using the data from Table \ref{tab_pm}.  Offsets are shown relative to \citet{Sadavoy19}.  Best-fit proper motions are given by the dashed lines.  Only the 1.3 mm (230 GHz) data are used for the fit.  Error bars include a 15 mas pointing uncertainty added in quadrature with the position uncertainty from Table \ref{tab_pm}.\label{fig_pm}}
\end{figure}

Figure \ref{fig_pm} shows the relative position of VLA 1623W compared to the epoch in \citet{Sadavoy19}.  We show the VLA~1623W results only as an example as the corresponding plot for VLA~1623B is very similar. The dashed line shows our adopted proper motion, which we measured using the data from \citet{Sadavoy19}, \citet{Harris18}, \citet{Murillo13}, and \citet{Chen13} only.  We excluded the data point from \citet{Maury12}, because they did not provide sufficient precision for a proper motion analysis \citep{Sadavoy18b}. We further excluded our new measurement presented here due to a noticeable offset of roughly 15 mas in RA and 30 mas in Dec compared to the general trend  (dashed line), which we found for both VLA~1623W and VLA~1623B.  As such, this 15 -- 30 mas offset appears to be systematic for the entire map.

A systematic offset in position could be explained by the present study using different calibrators.  Table \ref{tab_pm} lists the phase calibrator for each observation.  Most studies used J1625-2527, but this work used J1650-2943.  From the ALMA Technical Handbook, we can expect a 1~$\sigma$ pointing accuracy of $\sim 15$ mas, which is of comparable order to the systematic offsets.  To address possible astrometry uncertainties from the calibrators themselves, we added in quadrature a 15 mas position error to the position errors from Table \ref{tab_pm} before fitting for the proper motion using linear least squares.

We find a proper motion for VLA~1623W of%
\begin{eqnarray}
\mu_\alpha\cos{\delta} &=& -12.4\pm5.1\ \mbox{mas\ yr$^{-1}$}\\
\mu_\delta &=& -25.6\pm6.3\ \mbox{mas\ yr$^{-1}$}
,\end{eqnarray}
which are slightly different from \citet{Harris18} but within errors.  \citet{Harris18} measured a proper motion of VLA~1623W using their data and those of \citet{Murillo13} only.  There is a possibility that the position of VLA~1623W may also have differences due to wavelength since VLA 1623W is viewed nearly edge-on and shows evidence of flaring \citep{Michel22}.  There is a slight difference in RA position between \citet{Harris18} at 870 \um\ and the 1.3 mm measurements from \citet{Murillo13} and \citet{Sadavoy19}, whereas their Dec positions are in good agreement.  VLA~1623W appears nearly vertical in Dec, which would make any temperature variations with disk scale height appear only in RA. Nevertheless, \citet{Murillo13} have a different phase calibrator, which could also induce an offset.  Therefore, we cannot conclude that the shift in RA from \citet{Harris18} is due to temperature stratification. 

We also re-measured the proper motion of VLA~1623B from \citet{Sadavoy18b} using our revised errors that include adopting 15 mas pointing uncertainties.  The revised proper motion for VLA~1623B is%
\begin{eqnarray}
\mu_\alpha\cos{\delta} &=& -7.8\pm2.9\ \mbox{mas\ yr$^{-1}$}\\
\mu_\delta &=& -29.0\pm3.4\ \mbox{mas\ yr$^{-1}$}
,\end{eqnarray}
which is nearly identical in RA and has a slight difference in Dec within errors compared to \citet{Sadavoy18b}.

\begin{figure}
\includegraphics[width=0.485\textwidth]{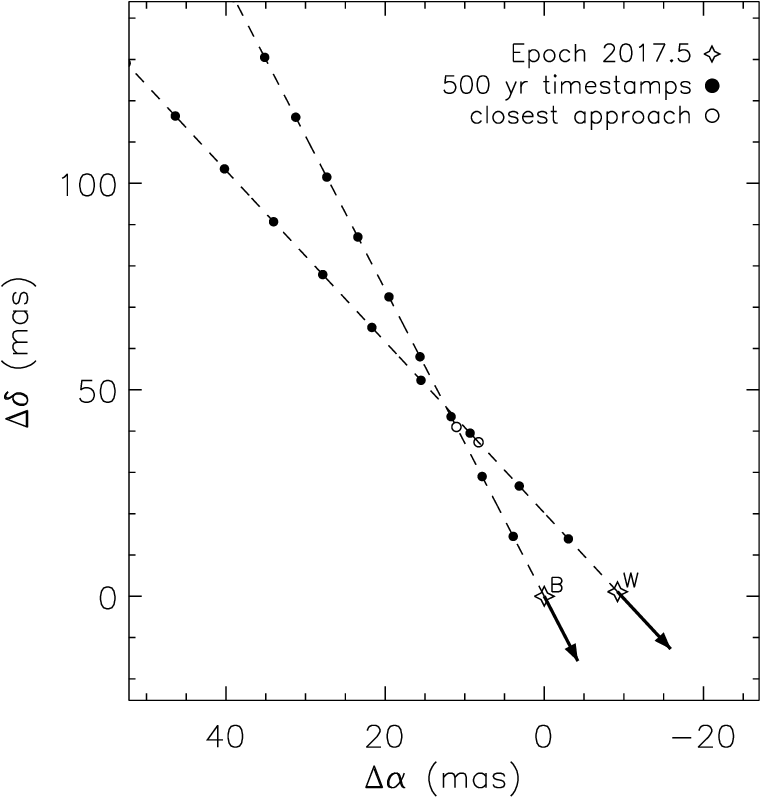}
\caption{Positions of VLA 1623B and VLA 1623W extrapolated in time based on their proper motion.  Arrows show the relative proper motions of both YSOs as measured here (for W) and in \citet{Sadavoy18b} (for B).  Solid circles show look back times for 5000 years in steps of 500 years.  The closest approach (in the plane of the sky) is represented by open circles.   \label{fig_pm_time}}
\end{figure}

We combined the above proper motions for VLA~1623W and VLA~1623B to determine when the two objects had their closest approach in the plane of the sky, assuming constant velocities.  Figure \ref{fig_pm_time} shows the extrapolation of both nominal proper motions with black circles showing look-back times in 500 yr intervals.  The closest approach is shown with open circles at a time of $\sim$ 1400 years, when the two sources were separated by  $\sim$ 4.6 arcsec ($\sim$ 650 au) in the plane of the sky. 

\begin{figure}
\includegraphics[width=0.485\textwidth]{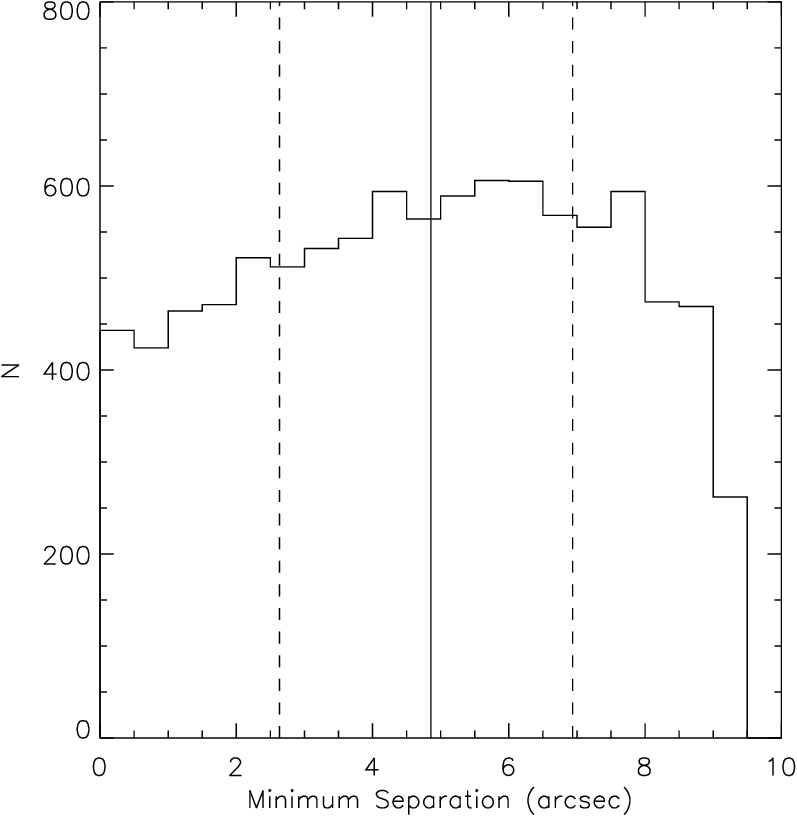} \includegraphics[width=0.485\textwidth]{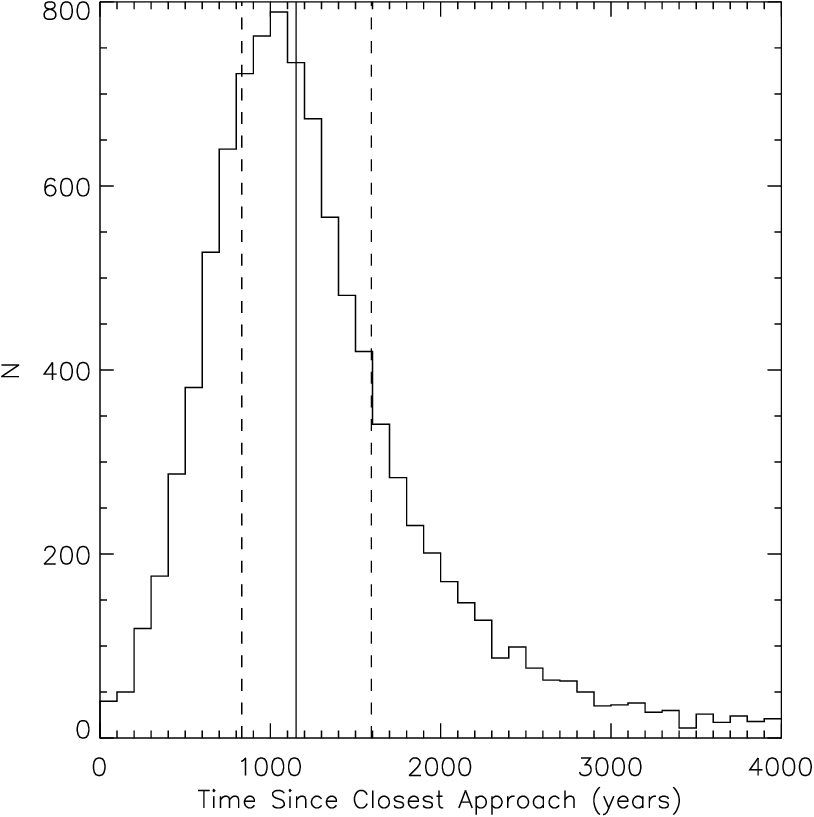}
\caption{Results for the minimum separation (top) and time of closest approach (bottom) for the Monte Carlo error analysis between the proper motions of VLA~1623B and VLA~1623W.  The solid lines show the median values and the dashed lines show the first quartile (25th percentile) and third quartile (75th percentile) for each distribution.  Note that the minimum separation has an upper limit of 9.3 arcsec, corresponding to the current separation between the two sources.  \label{fig_errhist_pm}}
\end{figure}

Nevertheless, the proper motions for both sources have large error bars ($> 10\%$), which make uncertainties on the closest approach difficult to constrain analytically.  We used a Monte Carlo error analysis for the proper motions for VLA~1623B and VLA~1623W to estimate the error in closest separation.  We assumed that the proper motion uncertainties given above correspond to the FWHM of a Gaussian distribution and then draw 10000 random values to add to the nominal proper motions.  For each new set of proper motions, we re-evaluated the closest approach. Figure \ref{fig_errhist_pm} shows histograms from this Monte Carlo analysis for the distribution of closest approach and the time since closest approach.  In both cases, we excluded roughly 200 data points for which the closest approach was given by the current epoch (e.g. in the case where the two sources are moving towards each other rather than away) as this case slightly skewed the statistics.  From the error analysis, we find a median minimum separation of 4.9 arcsec (first and third quartiles are 2.7 arcsec and 6.6 arcsec) and a median time of 1150 years (first and third quartiles are 830 years and 1590 years).   These results are in good agreement with the nominal case using the best-fit proper motions (without the errors).  Moreover, this analysis indicates that  VLA~1623W was unlikely to have a very close encounter as a plane of sky separation of $<$ 100 au is only at the 6th percentile.

\end{appendix}

\end{document}